\documentclass[prb,twocolumn,amsmath,amssymb,floatfix,footinbib,bibnotes,longbibliography]{revtex4-1}
\pdfoutput=1

\usepackage[colorlinks=true,citecolor=blue,linkcolor=magenta]{hyperref}
\usepackage{amsmath}
\usepackage{graphicx}
\usepackage{changepage}
\usepackage{fancyhdr}
\usepackage{amsthm, amssymb}
\usepackage{floatflt}
\usepackage{float}
\usepackage{array}
\usepackage[usenames,dvipsnames]{color}
\usepackage{mathtools}
\usepackage[normalem]{ulem}

\DeclareMathOperator{\Tr}{Tr}

\def \mr{\mathrm}

\def \GM{\boldsymbol{\mathrm{G}}}
\def \GMI{\boldsymbol{\mathrm{G^{-1}}}}
\def \bi{\boldsymbol{i}}
\def \IM{\boldsymbol{\mathrm{I}}}
\def \TM{\boldsymbol{\mathrm{T}}}
\def \bj{\boldsymbol{j}}
\def \br{\boldsymbol{r}}

\def \Nt{N_{\tau}}

\newcommand{\beq}{\begin{eqnarray}}
\newcommand{\eeq}{\end{eqnarray}}

\newcommand{\cd}{c^{\dagger}}
\newcommand{\cb}{\bar{c}}
\newcommand{\sr}{S_A^{(2)} }

\def \titlename {Dynamical mean-field theory for R\'{e}nyi entanglement entropy and mutual Information in Hubbard Model}
\def \authornames{Surajit Bera$^1$, Arijit Haldar$^{2,3}$ and Sumilan Banerjee$^1$}
\def \affiliations{$^1$Centre for Condensed Matter Theory, Department of Physics, Indian Institute 
	of Science, Bangalore 560012, India\\
	$^2$ S. N. Bose National Centre for Basic Sciences
JD Block, Sector-III, Salt Lake City, Kolkata - 700 106, India\\
	$^3$Department of Physics, University of Toronto, 60 St. George Street, Toronto, Ontario, M5S 1A7, Canada}

\begin{document}
	
	\title{\titlename}
	\author{\authornames}
	\affiliation{\affiliations}
	\email{surajit@iisc.ac.in}
	\email{arijit.haldar@bose.res.in}
	\email{sumilan@iisc.ac.in}
	\date\today

\begin{abstract}
Quantum entanglement, lacking any classical counterpart, provides a fundamental new route to characterize the quantum nature of many-body states. In this work, we discuss an implementation of a new path integral method [Phys.~Rev.~Res.~{\bf 2}, 033505 (2020)] for fermions to compute entanglement for extended subsystems in the Hubbard model within dynamical mean field theory (DMFT) in one and two dimensions. The new path integral formulation measures entanglement by applying a ``kick" to the underlying interacting fermions. We show that the R\'{e}nyi entanglement entropy can be extracted efficiently within the DMFT framework by integrating over the strength of the kick term. Using this method, we compute the second R\'{e}nyi entropy as a function of subsystem size for metallic and Mott insulating phases of the Hubbard model. We explore the thermal entropy to entanglement crossover in the subsystem R\'{e}nyi entropy in the correlated metallic phase. We show that the subsystem-size scaling of second R\'{e}nyi entropy is well described by the crossover formula which interpolates between the volume-law thermal R\'{e}nyi entropy and the universal boundary-law R\'{e}nyi entanglement entropy with logarithmic violation, as predicted by conformal field theory. We also study the mutual information across the Mott metal-insulator transition.

\end{abstract}

\maketitle 	
\section{Introduction} \label{sec:Intro}
Entanglement, arguably the strangest aspect of quantum mechanics, signifies the existence of true non-local quantum correlations. As a result, it has found enormous applications for characterizing quantum many-body states in condensed matter and high-energy physics, and as a resource for quantum computation \cite{Nielsen2000}. In condensed matter systems, entanglement can be used to distinguish various kinds of symmetry-broken and topological states, gapped or gapless phases \cite{Laflorencie} etc. For instance, entanglement provides an unambiguous indicator of topological order \cite{Wen,Kitaev} in quantum ground states. Entanglement has also emerged as an important measure for distinguishing high-energy states as well as non-equilibrium dynamics. For example, entanglement can be used to classify dynamical phases of isolated quantum systems as ergodic or many-body localized \cite{Nandkishore,Ehud,EhudDima}. 


Entanglement of a quantum system is quantified in terms of various measures, e.g., von Neumann and R\'{e}nyi entanglement entropies, mutual information and entanglement negativity \cite{Laflorencie,Cardy,Vidal2002}. These measures can be calculated by partitioning the overall system into two subsystems and computing the reduced density matrix of one of the subsystems by tracing over the other. To this end, the dependence of entanglement entropy on the size and geometry of the subsystem under various partitioning of the system are used to classify quantum many-body states and their non-local entanglement properties. For example, ground-states of gapped bosonic and fermionic systems in $d$ dimensions follow the so-called `area law' or `boundary-law' for entanglement entropy ($\sim L^{d-1}$) of a subsystem with length $L$ \cite{Laflorencie,RMP,Casini,Cardy}. In contrast, critical states in one dimension (1d) and fermionic systems with Fermi surface, i.e., standard metals, in any dimension exhibit a logarithmic violation \cite{Laflorencie,Korepin,Klich,Li2006,Swingle_PRL,Swingle2012,Swingle2012a,Senthil} of the area law, namely the subsystem entanglement entropy scales as $ L^{d-1}\ln{L}$. These characterizations of the many-body ground states 
are mainly obtained through powerful analytical results based on conformal field theory (CFT) methods \cite{Cardy,Calabrese_2009} and related arguments \cite{Swingle_PRL,Swingle2012,Swingle2012a,Senthil}, as well as numerical results for non-interacting systems \cite{Li2006,Casini}. For the latter, entanglement measures can be computed efficiently using the correlation matrix of the subsystem \cite{Casini}. However, numerical computations of entanglement entropy is much more challenging for interacting systems, typically limited to small systems accessible via exact diagonalization (ED) or 1d systems through density matrix renormalization group (DMRG) or heavily numerical and sophisticated quantum Monte Carlo (QMC) techniques \cite{Melko,Humeniuk2012,Grover,Assaad2014,Trebst,Troyer2014,Assaad2015,DEmidio2020}.

The above numerical methods have provided many useful insights into entanglement characteristics of interacting systems. However, there is a lack of complementary quantum many-body methods, e.g., mean-field theories, perturbation expansions, and other approximations, for computing entanglement entropy of interacting systems, unlike those for usual thermodynamic, spectroscopic and transport properties. The CFT techniques employ a replica path integral approach \cite{Cardy,Calabrese_2009} where bosonic and fermionic fields are defined on a non-trivial space-time manifold with complicated boundary conditions. The latter are often hard to implement within the standard quantum many-body methodology. To circumvent this difficulty, a new path integral approach was first developed in ref.\onlinecite{Sensarma} for bosons and was subsequently extended to fermionic systems \cite{Haldar,Moitra,AhanaLargeN}. In particular, ref.\onlinecite{Haldar} employed this method to compute R\'{e}nyi entanglement entropy of Fermi and non-Fermi liquid states of strongly interacting fermions described by Sachdev-Ye-Kitaev (SYK) and related models. The new method \cite{Haldar} replaces the complicated boundary conditions in the replica field theory for entanglement \cite{Cardy} by a fermionic self-energy that acts as a non-equilibrium kick. Using this new path integral formalism, here we develop a dynamical mean-field theory (DMFT) for R\`{e}nyi entanglement entropy in the paradigmatic Hubbard model \cite{Hubbard1963,FazekasBook,AuerbachBook} of strongly correlated electrons.

In the last three decades, single-site DMFT approximation and its cluster extensions \cite{Georges,Kotliar}, have gained popularity as a very successful approach to describe Mott metal-insulator transition and other associated electronic strong correlation phenomena, both in and out-of-equilibrium \cite{RMPNonEq}. The integration of DMFT with first-principle electronic structure methods, like density functional theory (DFT), has provided a viable route to compute and predict properties of strongly correlated materials \cite{Kotliar}. In the DMFT formulation, the strongly correlated lattice problem is reduced to a problem of a single impurity or cluster of sites coupled to a self-consistent bath \cite{Georges}. The original single-site implementation of DMFT neglects spatial correlations but captures non-trivial local dynamical quantum correlations and becomes exact in infinite dimension $d\to \infty$ \cite{Georges}. The later cluster extensions of DMFT \cite{Hettler2000,Kotliar2001,Jarrel,Kotliar}, along with state-of-the-art impurity solver like continuous-time quantum Monte Carlo (CTQMC), incorporates back some of the spatial correlations, and can even provide a good description of the properties of one-dimensional systems \cite{Bolech2003,Capone2004,Kotliar}. 

 In this work, we compute entanglement properties of the correlated metallic and insulating phases across the Mott metal-insulator transition in the Hubbard model. We use the cavity method \cite{Georges} for the entanglement path integral of ref.\onlinecite{Haldar} to derive single-site DMFT self-consistency equations for obtaining the second R\'{e}nyi entropy $S_A^{(2)}$ of a contiguous subsystem $A$. We show that the R\'{e}nyi entropy can be extracted by integrating over the strength of a non-equilibrium `kick' perturbation acting on the imaginary-time evolution. Remarkably, this only requires the knowledge of onsite single-particle Green's function for the subsystem, even in the interacting system, albeit in the presence of the kick. Due to the entanglement cut(s) and the non-equilibrium kick, both the lattice and time translation symmetry are broken in the entanglement path integral. Thus the single-site DMFT is implemented as an inhomogeneous non-equilibrium DMFT. To this end, we develop an efficient recursive Green's function method to solve the DMFT self-consistency equations. Given the computational complexity of the problem, we only consider the Hubbard model at half filling and employ a simple DMFT impurity solver, namely the iterative perturbation theory (IPT). The latter is known to work very well when compared to more accurate exact diagonalization and QMC impurity solvers for the half-filled Hubbard model in equilibrium \cite{Georges}. Our DMFT formulation is general and can be extended in future to incorporate cluster generalizations of DMFT \cite{Hettler2000,Kotliar2001,Jarrel,Kotliar} and more accurate impurity solvers, e.g., CTQMC\cite{Gull}.
 
 Using the inhomogeneous non-equilibrium single-site DMFT for subsystem R\'{e}nyi entropy, we compute $S_A^{(2)}$ in the Hubbard model as a function of temperature $T$, interaction $U$, and linear size of the subsystem $N_A$ in 1d, and for 2d cylindrical geometry. In particular, we ask how the entanglement properties of correlated metal described by completely local self-energy approximation within single-site DMFT compare with those expected from CFT \cite{Cardy,Korepin,Calabrese_2009} and related arguments \cite{Swingle_PRL,Swingle2012,Swingle2012a,Senthil}. At high temperature, subsystem R\'{e}nyi entropy $S_A^{(2)}$ is dominated by thermal entropy and, at low temperature, by entanglement \cite{Korepin,Cardy,Calabrese_2009,Ryu2006,Senthil}. We indeed find a crossover from thermal to entanglement behavior in $S_A^{(2)}$. Specifically, we show that this crossover in DMFT metallic state is well described by the known CFT crossover formula in 1d and its higher dimensional generalization \cite{Korepin,Cardy,Ryu2006}. We also compare our results for $S_A^{(2)}(N_A,T)$ with that available from  QMC \cite{Trebst}. 
 
 As a measure of entanglement at finite temperature \cite{Wolf2008}, we extract R\'{e}nyi mutual information between the subsystem $A$ and the rest of the system from $S_A^{(2)}(N_A)$. We find that the mutual information has a hysteresis across the first-order Mott metal-insulator transition \cite{Georges} in the $U-T$ plane, culminating at the critical point where the transition becomes second order. There have been previous studies \cite{Motome,Tremblay,Walsh2019,Walsh2020} of von-Neumann entropy, mutual information as well as entanglement spectrum of a single site, or a few sites within a cluster, via cellular DMFT (CDMFT).
 Such \emph{local} entanglement measures can be computed within the usual equilibrium DMFT formulation. However, the full subsystem size dependence of the entanglement entropy and mutual information cannot be obtained through such equilibrium DMFT. On the contrary, the general method we develop here can be applied for extended subsystems of arbitrary size and shape and requires an entirely different implementation through the new path integral technique \cite{Haldar} and non-equilibrium kick term.
 
 The rest of the paper is organized as follows. In Sec.\ref{sec:PathIntegral}, we briefly revise the general path integral formalism of ref.\onlinecite{Haldar} for the second R\'{e}nyi entropy and discuss how the R\'{e}nyi entropy can be extracted by integrating over the strength of a non-equilibrium kick term. The DMFT approximation for the entanglement path integral is discussed in Sec.\ref{sec:DMFTEntanglement} in the context of half-filled Hubbard model. The numerical solution of the DMFT self-consistency equations and some of the benchmarks through the comparisons with the non-interacting limit and previous QMC simulations are discussed in Sec.\ref{sec:NumBenchmark}. We then discuss our main results for the subsystem size dependence of R\'{e}nyi entropy in 1d and 2d Hubbard model and the thermal entropy to entanglement crossover in Secs.\ref{sec:1DHubbard},\ref{sec:2DHubbard}. In Sec.\ref{sec:MutualInfo}, we discuss the mutual information across the Mott metal-insulator transition. We summarize our results and discuss possible future scopes and extensions of our work in Sec.\ref{sec:Conclusion}. The details of the numerical implementations of the DMFT equations, benchmarks and analysis of the results are given in the Appendices \ref{secsupp:DMFT_A}-\ref{secsupp:DMFT_2d}.

\section{The path integral for subsystem R\'{e}nyi entropy}\label{sec:PathIntegral}
In this section, we briefly discuss the path integral formalism~\cite{Haldar} for subsystem R\'{e}nyi entropy of fermions in a thermal state. For concreteness, we consider a system of spin-1/2 fermions on a lattice with $N$ sites in thermal state at a temperature $T$ described by a density matrix $\rho=e^{-\beta\mathcal{H}}/Z$ with Hamiltonian $\mathcal{H}$. Here $\beta=1/T$ ($k_\mathrm{B}=1$) and $Z$ is the partition function. We obtain the reduced density matrix for subsystem $A$ with $N_A$ site by integrating the degrees of freedom of the rest of the system $B$, i.e., $\rho_A={\rm Tr_ B}(\rho)$. Then, the $n$-th R\'{e}nyi entropy for the subsystem $A$ is 
\begin{align}
    S^{(n)}_A = \frac{1}{1-n}\ln {\rm Tr_A}\big[\rho^n_A\big]. 
\end{align}
In this work, we only consider the second R\'{e}nyi entropy $S_A^{(2)}$ for simplicity. The method can be generalized easily for higher-order R\'{e}nyi entropies.
The path integral is constructed based on the following identity~\cite{Haldar}
\begin{align}
	e^{-S_A^{(2)}}\equiv \Tr_A[\rho_A^2] = \int d^2\xi f(\xi)\Tr[\rho_AD(\xi_1)]\Tr[\rho_AD(\xi_2)], \label{eq:TraceIdentity}
\end{align}
 where `$\mathrm{Tr}$' denotes trace over the entire system, and $D(\xi_\alpha)=\exp{(\sum_{i\in A,\sigma}c_{i\sigma}^\dagger\xi_{i\sigma\alpha})}\exp{(-\sum_{i\in A,\sigma}\bar{\xi}_{i\sigma\alpha}c_{i\sigma}^\dagger)}$ ($\alpha=1,2$) is normal-ordered \emph{fermionic} displacement operator~\cite{Glauber} written in terms of fermionic creation and annihilation operators $ c_{i\sigma}^\dagger,\ c_{i\sigma}$ at site $i$, in the $A$ subsystem, with spin $\sigma=\uparrow,\downarrow$. Also, $\bar{\xi}_{i\sigma\alpha},\ \xi_{i\sigma\alpha}$ are static auxiliary Grassmann fields  in $A$ with $d^2\xi=\prod_{i\in A,\sigma\alpha} d\bar{\xi}_{i\sigma\alpha}d\xi_{i\sigma\alpha}$, and 
 \begin{align}
f(\xi)=2^{N_A}e^{-\frac{1}{2}\sum_{i\in A, \sigma}(\bar{\xi}_{i\sigma 1}\xi_{i\sigma 1}+\bar{\xi}_{i\sigma 2}\xi_{i\sigma 2}-\bar{\xi}_{i\sigma 1}\xi_{i\sigma 2}+\bar{\xi}_{i\sigma 2}\xi_{i\sigma 1})}
\end{align}
is a Gaussian factor. The term $\mathrm{Tr}[\rho_AD(\xi)]$, called the characteristic function, can be written in terms of a coherent-state path integral~\cite{Haldar},
\begin{align}
	e^{-S_A^{(2)}} =\frac{1}{Z^2}\int d^2\xi f(\xi)\mathcal{D}(\bar{c},c)e^{-(\mathcal{S}+\mathcal{S}_{\xi})} \label{eq:RenyiSource}
\end{align}
where $\mathcal{S}=\int_0^\beta[\sum_{i\sigma\alpha}\bar{c}_{i\sigma\alpha}(\tau)(\partial_\tau+\mu)c_{i\sigma\alpha}(\tau)+\mathcal{H}(\bar{c},c)]$, and $\mathcal{S}_{\xi}=\int_0^\beta \sum_{i\in A,\sigma}[\bar{c}_{i\sigma\alpha}(\tau)\delta(\tau-\tau_0^+)\xi_{i\sigma\alpha}-\bar{\xi}_{i\sigma\alpha}\delta(\tau-\tau_0)c_{i\sigma\alpha}(\tau)]$ is a source term which acts on the $A$ subsystem at imaginary time $\tau_0$ and originates from the displacement operator in Eq.\eqref{eq:TraceIdentity}. The imaginary time $\tau_0$ is arbitrary and can be placed anywhere on the thermal cycle $0\leq \tau<\beta$. The fermionic fields have the usual anti-periodic boundary condition $c(\tau+\beta)=-c(\tau)$. For non-interacting systems, it is straightforward to integrate out \cite{Sensarma,Haldar, Moitra} the fermionic fields $\bar{c},c$ and the auxiliary fields $\bar{\xi},\xi$ to obtain the R\'{e}nyi entropy. As discussed in ref.\onlinecite{Haldar}, for interacting systems treated within some non-perturbative approximations, like in large-$N$ models, it is advantageous to first integrate out the Gaussian auxiliary fields in Eq.\eqref{eq:TraceIdentity} and obtain 
\begin{align}
    e^{-S_A^{(2)}}\equiv \frac{Z_A^{(2)}}{Z^2} =\frac{1}{Z^2}\int \mathcal{D}(\bar{c},c)e^{-(\mathcal{S}+\mathcal{S}_\mr{kick})} \label{eq:RenyiKick},
\end{align}
where
\begin{align} 
\mathcal{S}_\mr{kick} = \sum_{i\in A, \alpha \beta \sigma}\cb_{i\sigma\alpha}(\tau)M_{\alpha\beta}\delta(\tau-\tau_0^{+})\delta(\tau'-\tau_0)c_{i\sigma\beta}(\tau'), \label{eq:KickAction}
\end{align}
henceforth referred as the \emph{kick} term, corresponds to an effective time-dependent self-energy for the fermions at $\tau_0$. The matrix 
 \beq \label{eq:M}
 M &=& \begin{bmatrix}
1 & 1\\
-1 & 1 
\end{bmatrix}\eeq 
couples the two replicas $\alpha=1,2$. In ref.\onlinecite{Haldar}, we have used the path integral representation of Eq.\eqref{eq:RenyiKick} to evaluate $S_A^{(2)}$ of the SYK model and its several extensions. Below we show that the same representation can be utilized to formulate a DMFT for R\'{e}nyi entanglement entropy in the Hubbard model. 

\subsection{Subsystem R\'{e}nyi entropy via integration of the kick term}
Using Eq.\eqref{eq:RenyiKick}, the 2nd R\'{e}nyi entropy $S_A^{(2)}$ can be formally written as
\begin{align}
S_A^{(2)}=\beta(\Omega^{(2)}_A-2\Omega),
\end{align}
where we define $\Omega_A^{(2)}\equiv-T\ln Z_A^{(2)}$ and $\Omega=-T\ln Z$ is the thermodynamic grand potential. However, direct computation of both $\Omega_A^{(2)}$ and $\Omega$ for interacting systems is difficult in general and typically requires thermodynamic or coupling constant integration \cite{Haldar,Fetter}. Here we find a new way to extract $S_A^{(2)}$ by using the kick term in Eq.\eqref{eq:RenyiKick}. We consider the following quantity
\begin{align}
e^{-S_A^{(2)}(\lambda)}=\frac{Z_A^{(2)}(\lambda)}{Z^2}=\frac{1}{Z^2}\int \mathcal{D}(\bar{c},c)e^{-(\mathcal{S}+\lambda \mathcal{S}_\mathrm{kick})}, \label{eq:RenyiKicklambda}
\end{align}
which reduces to $S_A^{(2)}(\lambda=1)=S_A^{(2)}$, the second R\'{e}nyi entropy, for $\lambda=1$, and $S_A^{(2)}(\lambda=0)=0$. In the above, by taking the derivative with respect to $\lambda$, we get
\begin{align}
 \partial_{\lambda}  \sr(\lambda) = \frac{\int \mathcal{D}(\cb, c)e^{-(\mathcal{S}+\lambda \mathcal{S}_{\rm kick}) } \mathcal{S}_{\rm kick} }{\int \mathcal{D}(\cb, c)e^{-(\mathcal{S}+\lambda \mathcal{S}_{\rm kick}) }}=\langle \mathcal{S}_\mr{kick}\rangle_{Z_A^{(2)}(\lambda)}, 
\end{align}
i.e., the expectation value of the kick term with respect to the effective partition function $Z_A^{(2)}(\lambda)$. Integrating the above equation over $\lambda$ from 0 to 1, we obtain an expression for the second R\'{e}nyi entropy
\begin{align}
    S_A^{(2)}=\int_0^1d\lambda \langle \mathcal{S}_\mr{kick}\rangle_{Z_A^{(2)}(\lambda)}. \label{eq:KickIntegration}
\end{align}
 The great advantage of the above expression is that the kick term is quadratic in Grassmann variables $\cb, c$. {In the above, we have assumed that no phase transition occurs as we vary $\lambda$. Such transition as a function of $\lambda$ in the entanglement action might be present and will be interesting to study in future. Assuming that there are no transitions with $\lambda$,} the R\'{e}nyi entropy can be obtained from Eq.\eqref{eq:KickIntegration} using
 \begin{align} \label{eq:Avg_Skick}
\langle \mathcal{S}_{\rm kick} \rangle_{Z_A^{(2)}(\lambda)} =  \sum_{i\in A,\alpha\beta\sigma} M_{\alpha\beta}G_{i\sigma\beta, i\sigma\alpha}(\tau_0, \tau^{+}_{0}),
\end{align}
where the imaginary-time local single-particle Green's function 
\beq 
G_{i\sigma\alpha,i\sigma\beta}(\tau,\tau')=-\langle \mathcal{T}_{\tau}c_{i\sigma\alpha}(\tau)\cb_{i\sigma\beta}(\tau')\rangle_{Z_A^{(2)}(\lambda)}. \label{eq:RenyiGreenFn}
\eeq 
The Green's function, however, needs to be evaluated in the presence of the kick term with variable $\lambda$. In the next section, we show how the Green's function can be obtained through the DMFT approximation. 

\section{Dynamical mean-field theory for the second R\'{e}nyi entropy in the Hubbard model} \label{sec:DMFTEntanglement}

We consider the nearest-neighbor Hubbard model 
\beq 
\mathcal{H} = \sum_{ij,\sigma}t_{ij}\cd_{i\sigma}c_{j\sigma} -\mu \sum_{i}n_i + U\sum_{i}n_{i\uparrow} n_{i\downarrow} \label{eq:HubbardModel}
\eeq 
where $t_{ij}$ is the hopping amplitude between lattice sites ($i=1,\dots,N$) on 1d and 2d square lattice, $\mu$ is the chemical potential, and $U$ is the onsite repulsive interaction strength between fermions with opposite spins $\sigma=\uparrow, \downarrow$. Here, $n_{i\sigma}=\cd_{i\sigma}c_{i\sigma}$ and  $n_i=\sum_{\sigma}n_{i\sigma}$ are the electronic number operators. We write down the entanglement action $\mathcal{S}_\lambda=\mathcal{S}+\lambda S_\mr{kick}$ of Eq.\eqref{eq:RenyiKicklambda} for the Hubbard model as


\begin{align} \label{eq:compact_S_ent}
\mathcal{S}_\lambda &= -\int_{0}^{\beta} d\tau d\tau' \sum_{ij,\sigma} \cb_{i\sigma\alpha}(\tau)G^{-1}_{0, i\alpha, j\beta}(\tau, \tau')c_{j\sigma \beta}(\tau') \notag \\
&+ \int_{0}^{\beta} d\tau U\sum_{i\alpha}n_{i\uparrow \alpha}(\tau)n_{i\downarrow \alpha}(\tau),
\end{align}
where $G^{-1}_{0,i\alpha, j\beta}(\tau,\tau')$ is inverse non-interacting lattice Green's function in the presence of entanglement cut(s) between $A$ and the rest of the systems, i.e.
\begin{align}
G^{-1}_{0, i\alpha, j\beta}(\tau, \tau')& =  -[(\partial_{\tau}-\mu)\delta_{ij}+t_{ij}]\delta(\tau- \tau')\delta_{\alpha, \beta} \notag\\
&-\lambda \delta_{i\in A}\delta_{ij}M_{\alpha,\beta} \delta(\tau-\tau_0^{+})\delta(\tau'-\tau_0).
\end{align}
As mentioned earlier in Sec.\ref{sec:Intro}, the self-energy kick, which only acts on $A$ ($\delta_{i\in A}=1$ for $i\in A$ and zero otherwise), breaks both lattice and time-translation symmetry in this formulation. As a result, we construct a single-site inhomogeneous non-equilibrium DMFT. We use the cavity method \cite{Georges} to reduce the lattice problem into effective single-site problems for each of the sites $i=1,\dots,N$, described by the generating functions
\beq 
Z^{(2)}_{\lambda,i} = \int \mathcal{D}(\cb, c) e^{-\mathcal{S}_{\lambda,i}} \label{eq:ImpurityProblem}
\eeq 
where the effective action $\mathcal{S}_{\lambda,i}$ is given by
\begin{align}
\mathcal{S}_{\lambda,i} &= -\int_{0}^{\beta} d\tau d\tau'   \sum_{\sigma\alpha\beta} \cb_{\sigma\alpha}(\tau)\mathcal{G}^{-1}_{i, \alpha\beta}(\tau, \tau')c_{\sigma \beta}(\tau') \notag\\
&+ \int_{0}^{\beta} d\tau U\sum_{\alpha}n_{\uparrow \alpha}(\tau)n_{\downarrow \alpha}(\tau) .
\end{align}
Here $\mathcal{G}_{i, \alpha\beta}(\tau,\tau')$ is the dynamical \emph{Weiss} field, such that
\begin{align} \label{eq:DMFTeq1}
\mathcal{G}^{-1}_{i}(\tau, \tau')&=   -(\partial_{\tau}-\mu)\delta(\tau- \tau')\mathcal{I}
- \Delta_{i}(\tau, \tau')\notag\\
&- \lambda \delta_{i\in A}M\delta(\tau-\tau_0^{+})\delta(\tau'-\tau_0),
\end{align}
is a $2\times 2$ matrix in the entanglement replica space, and $\mathcal{I}$ is the identity matrix in the same space. In the above, we have also assumed a paramagnetic state. Of course, like in equilibrium DMFT~\cite{Georges}, the formulation can be easily extended, to describe entanglement in ordered states, such as the antiferromagnetic N\'{e}el state in the Hubbard model. The matrix $\Delta_i(\tau,\tau')$ in Eq.\eqref{eq:DMFTeq1} is the hybridization function which can be expressed in terms of the lattice Green's function as discussed below. The impurity Green's function is related to the Wiess field via the Dyson equation,
\beq \label{eq:DMFTeq2}
G^{-1}_i(\tau,\tau') = \mathcal{G}^{-1}_i(\tau,\tau') - \Sigma_i(\tau,\tau'),
\eeq 
where $\Sigma_i(\tau,\tau')$ is the impurity self-energy. The Green's function can be obtained by solving the impurity problem using some approximate or exact impurity solvers \cite{Georges,RMPNonEq}, e.g. CTQMC \cite{Gull}. In this work, for simplicity, and as a first attempt to compute entanglement via DMFT within the new formalism \cite{Haldar}, we use iterative perturbation theory (IPT) \cite{Georges} to obtain the self-energy in Eq.\eqref{eq:DMFTeq2}. 

We consider the particle-hole symmetric half-filling case with the chemical potential $\mu=U/2$. At half-filling, IPT, which essentially retains the self-energy up to second-order in $U$, is known to work very well \cite{Georges} in equilibrium, especially in the metallic phase. As well known, in this case, IPT coincides with the exact result for both $U\to 0$ and $U\to \infty$, i.e., the atomic limit, and thus it interpolates well between the two limits even at intermediate $U$. For the effective non-equilibrium problem [Eq.\eqref{eq:ImpurityProblem}], we also use the IPT as an approximate impurity solver. The IPT self-energy in our case is given by 
\begin{align} \label{eq:DMFTeq3}
\Sigma_{i,\alpha\beta}(\tau,\tau') &= U G_{{ii},{\alpha\beta}}(\tau,\tau^{+})\delta(\tau'-\tau^+)\delta_{\alpha\beta} \nonumber \\
&- U^2\tilde{\mathcal{G}}_{i,\beta\alpha}^2(\tau,\tau') \tilde{\mathcal{G}}_{i,\alpha\beta}(\tau',\tau).
\end{align}
Here the first term is Hartree self-energy, and the second one is the second-order self-energy obtained using Hartree corrected Green’s function
\begin{align}
\tilde{\mathcal{G}}^{-1}_{i,\alpha\beta}(\tau,\tau') &=& \mathcal{G}^{-1}_{i,\alpha\beta}(\tau,\tau')-U G_{{ii},{\alpha\beta}}(\tau,\tau^{+})\delta(\tau'-\tau^+)\delta_{\alpha\beta}.
\end{align}
 Within the single-site DMFT approximation, we assume the self-energy in the lattice problem to be local and the same as the impurity self-energy. Thus the lattice Green’s function $G_{i\alpha,j\beta}(\tau,\tau')$ is obtained from the lattice Dyson equation 
\begin{align} \label{eq:LatticeGreenEq}
\int_{0}^{\beta}&d\tau'' \sum_{k\gamma} \big[G^{-1}_{0, i\alpha, k\gamma}(\tau, \tau'')-\delta_{ik}\Sigma_{i,\alpha\gamma}(\tau,\tau'')]G_{k\gamma,j\beta}(\tau'', \tau') 
\notag \\
&= \delta_{ij}\delta_{\alpha\beta}\delta(\tau-\tau')
\end{align}
The DMFT loop is closed by relating the lattice Green’s function with the hybridization function $\Delta_i(\tau,\tau')$. The latter can be obtained via the cavity method (see Appendix-\ref{RecursiveGcavity}) in terms of the cavity Green's function as
\beq \label{eq:hybridization_eq1}
\Delta_{i,\alpha\beta}(\tau,\tau') = \sum_{jl} t_{ij}t_{il}G^{(i)}_{j\alpha,l\beta}(\tau,\tau'),
\eeq
where the cavity Green's function, obtained with the $i$-th site removed from the original lattice, is related to full lattice Green's function via
\begin{align} \label{eq:cavity_eq1}
G^{(i)}_{j\alpha,l\beta}(\tau,\tau') &= G_{j\alpha, l\beta}(\tau,\tau')
\notag\\
-\int d\tau_1 d\tau_2\sum_{\gamma\delta} &G_{j\alpha,i\gamma}(\tau,\tau_1) [G_{i\gamma, i\delta}(\tau_1,\tau_2)]^{-1} G_{i\delta,l\beta}(\tau_2,\tau').
\end{align}
The above closes the DMFT self-consistency loop. 
For our numerical computations, we further make the large-connectivity Bethe lattice approximation \cite{Georges} for the Cavity Green's function. As a result, for the model [Eq.\eqref{eq:HubbardModel}] with only nearest-neighbor hopping, Eq.\eqref{eq:hybridization_eq1} becomes
\beq \label{eq:BetheApprox}
\Delta_{i,\alpha\beta}(\tau,\tau') = t^2\sum_{j}' G_{j\alpha,j\beta}(\tau,\tau'),
\eeq
where $\sum_j'$ indicates that the summation is over only the nearest-neighbors of $i$. This approximation makes the computation easier keeping the essential features of the finite dimensionality through the lattice Green's function. 

Computationally, the most expensive part of the DMFT loop here is the inversion of Eq.\eqref{eq:LatticeGreenEq} to obtain the lattice Green's function $\GM$, a matrix in indices $(i\alpha\tau, j\beta\tau')$. As discussed in the next section and in Appendix \ref{secsupp:DMFT_A}, we discretize the imaginary time and use a recursive Green's  function method for large systems to obtain $\GM$. We also benchmark our results by doing direct inverse in Eq.\eqref{eq:LatticeGreenEq} for small systems.


\section{Numerical solution of DMFT equations to obtain $\sr$} \label{sec:NumBenchmark}
We solve the DMFT self-consistency Eqs.(\ref{eq:DMFTeq1}, \ref{eq:DMFTeq2}, \ref{eq:DMFTeq3}, \ref{eq:LatticeGreenEq}, \ref{eq:hybridization_eq1}, \ref{eq:BetheApprox}) by discretizing them in imaginary time $\tau$ with discretization step $\delta \tau$ as discussed in Appendix \ref{secsupp:DMFT_A} in detail. To evaluate $S_A^{(2)}$ from Eq.\eqref{eq:KickIntegration}, we perform the DMFT for a range of values of $\lambda$ between 0 to 1 with step $\delta \lambda$ as discussed in Appendix \ref{secsupp:kickterm}.  After obtaining the DMFT self-consistent Green's function [Eq.\eqref{eq:RenyiGreenFn}] for different $\lambda$, we use Eqns.\eqref{eq:Avg_Skick} and \eqref{eq:KickIntegration} to compute $\sr(\delta \tau)$ for a given discretization $\delta \tau$. We compute $\sr(\delta\tau)$ for different values of $\delta \tau$ and extrapolate to $\delta \tau \to 0$ limit to finally obtain $\sr$. In Appendix \ref{secsupp:extrap}, we discuss the extrapolation of $\sr(\delta\tau)$ to $\delta\tau \to 0$.
As mentioned in the preceding section, we employ a recursive Green's function method to obtain the lattice Green's function from Eq.\eqref{eq:LatticeGreenEq} (see Appendix \ref{RecursiveG}). By using the recursive Green's function method we can compute $S_A^{(2)}$ for reasonably large systems, $N\leq 100$ in 1d, and $N\leq 20\times 20$ in 2d up to low temperatures ($T\geq 0.05$, in units of nearest-neighbor hopping amplitude $t$). The results reported in the main text are for periodic boundary condition (PBC). We also discuss some results for open boundary condition (OBC) in Appendix \ref{secsupp:OBC}.

\begin{figure}[ht]
    \centering
\includegraphics[width=0.9\linewidth]{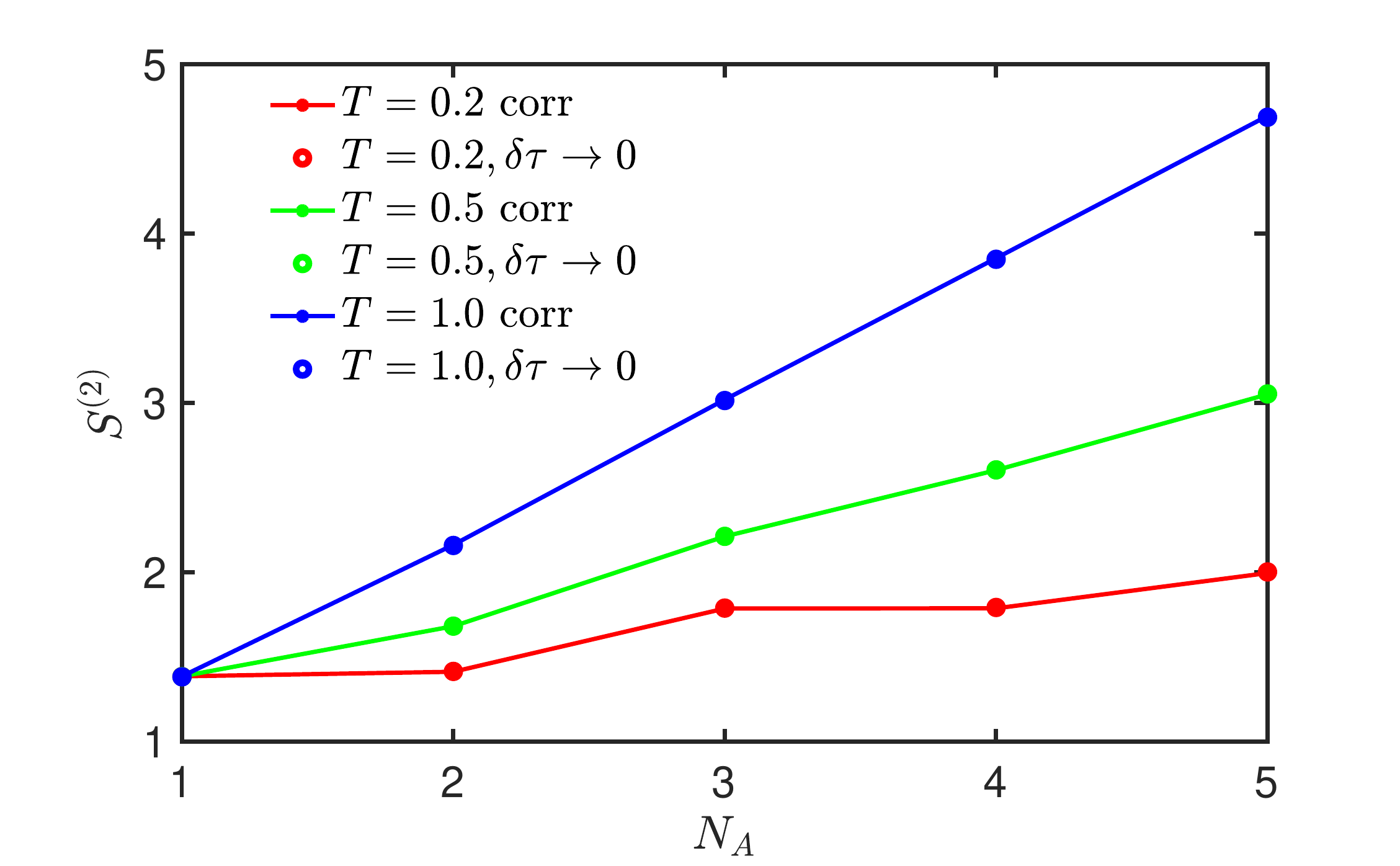}
\caption{The extrapolated $\sr(\delta\tau\to 0)$ (open circles) for $U=0$ is compared with the $\sr$ from correlation matrix calculation (closed circle+line, `corr') for $N=10$ and three different temperatures $T=0.2, 0.5, 1.0$. }
\label{supp_fig:benchmark_NI}
\end{figure} 

\subsection{Comparison with the non-interacting limit and QMC} \label{subsec:benchmark}
To benchmark the kick integration method of Eq.\eqref{eq:KickIntegration} and the extrapolation $\sr(\delta\tau\to 0)$, we first compare the results for $S_A^{(2)}$ for the non-interacting case ($U=0$) with that calculated directly using the correlation matrix $C_{ij}=\mathrm{Tr}[\rho c_i^\dagger c_j]$ for $i,j\in A$. The correlation matrix can be easily evaluated using the single-particle eigenenergies and eigenfunctions of the tight-binding model of Eq.\eqref{eq:HubbardModel} for $U=0$. The second R\'{e}nyi entropy is obtained from $S_A^{(2)}=-\mathrm{Tr}\ln[(1-C)^2+C^2]$ \cite{Casini,Haldar}. As shown in Fig.\ref{supp_fig:benchmark_NI}, $\sr(\delta\tau\to 0)$ for different temperatures matches very well with the corresponding $\sr$ from correlation matrix calculation for a non-interacting system of size $N=10$.

\begin{figure}[ht]
    \centering
\includegraphics[width=0.95\linewidth]{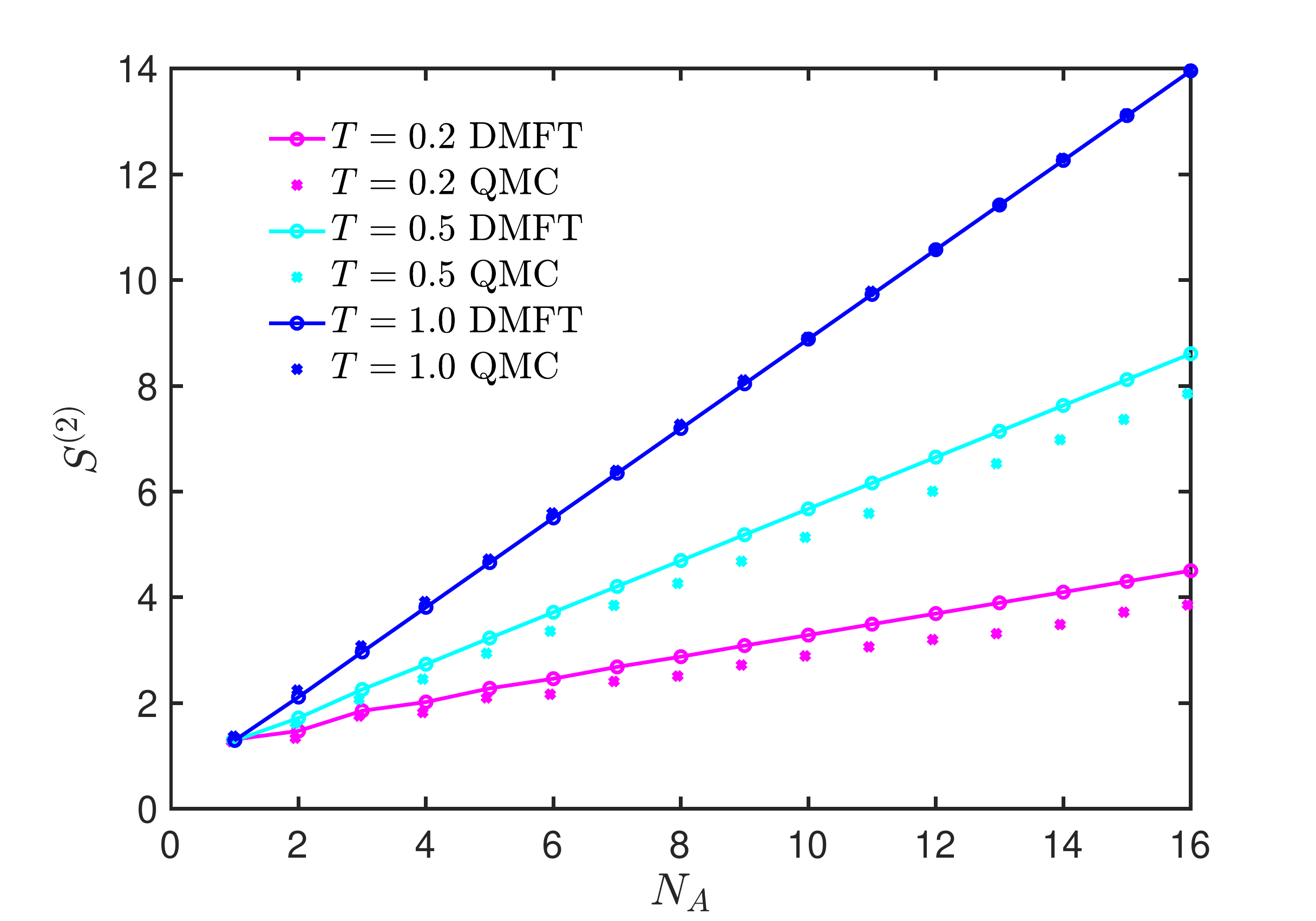} 
\caption{DMFT results for $S_A^{(2)}$ as a function of subsystem size $N_A$ in 1d Hubard model are compared with QMC data from ref.\onlinecite{Trebst} at three temperatures $T=1.0,0.5,0.2$ for system size $N=32$ and $U=2$.}
\label{mainfig_QMC_vs_DMFT}
\end{figure}

We next compare our DMFT results for $\sr$ with QMC data taken from ref.\onlinecite{Trebst} for $N=32$ and $U=2.0$. In Fig.(\ref{mainfig_QMC_vs_DMFT}), the $\sr$ as a function of subsystem size $N_A$ for this two method is shown for high ($T=1.0$) to intermediate temperatures ($T=0.5, 0.2$). In high $T$, the $\sr$ the DMFT and QMC results coincide. Even at intermediate temperatures, the comparison is reasonable given the fact that 1d is the worst-case scenario for a mean-field approach like single-site DMFT, and that too, employing an approximate impurity solver like IPT. Nevertheless, it has been shown \cite{Bolech2003,Capone2004,Kyung2006} that cluster extension of DMFT can capture some of the subtle Luttinger-liquid physics in 1d arising from long-distance correlations. Hence, cluster extensions of our DMFT approach will be able to provide in the future a good description of entanglement properties even in 1d. 

Given the above benchmarks, in the next sections, we study the subsystem-size dependence and the entropy to entanglement crossover of $S_A^{(2)}$, first in 1d, and then for the 2d Hubbard model. 


\section{$\sr$ in 1d Hubbard model} \label{sec:1DHubbard}
In 1d, single-site DMFT gives rise to a metal-insulator transition at finite $U$ \cite{Bolech2003,Kyung2006} at half filling, unlike the exact Bethe ansatz solution \cite{Lieb1968}. The latter leads to a metallic state only at $U=0$ and gapped states for any $U>0$. The metallic state in DMFT is a relic of the infinite dimension inherent in the local self-energy approximation in single-site DMFT, even though some  effects of finite dimension are fed back through the lattice self-consistency. We first look into $S_A^{(2)}(N_A,T)$ of this mean-field metallic state in 1d since the DMFT for entanglement formulated in Sec.\ref{sec:DMFTEntanglement} is numerically much easier to implement in 1d than in higher dimensions.

\begin{figure}[ht]
    \centering
\includegraphics[width=0.95\linewidth]{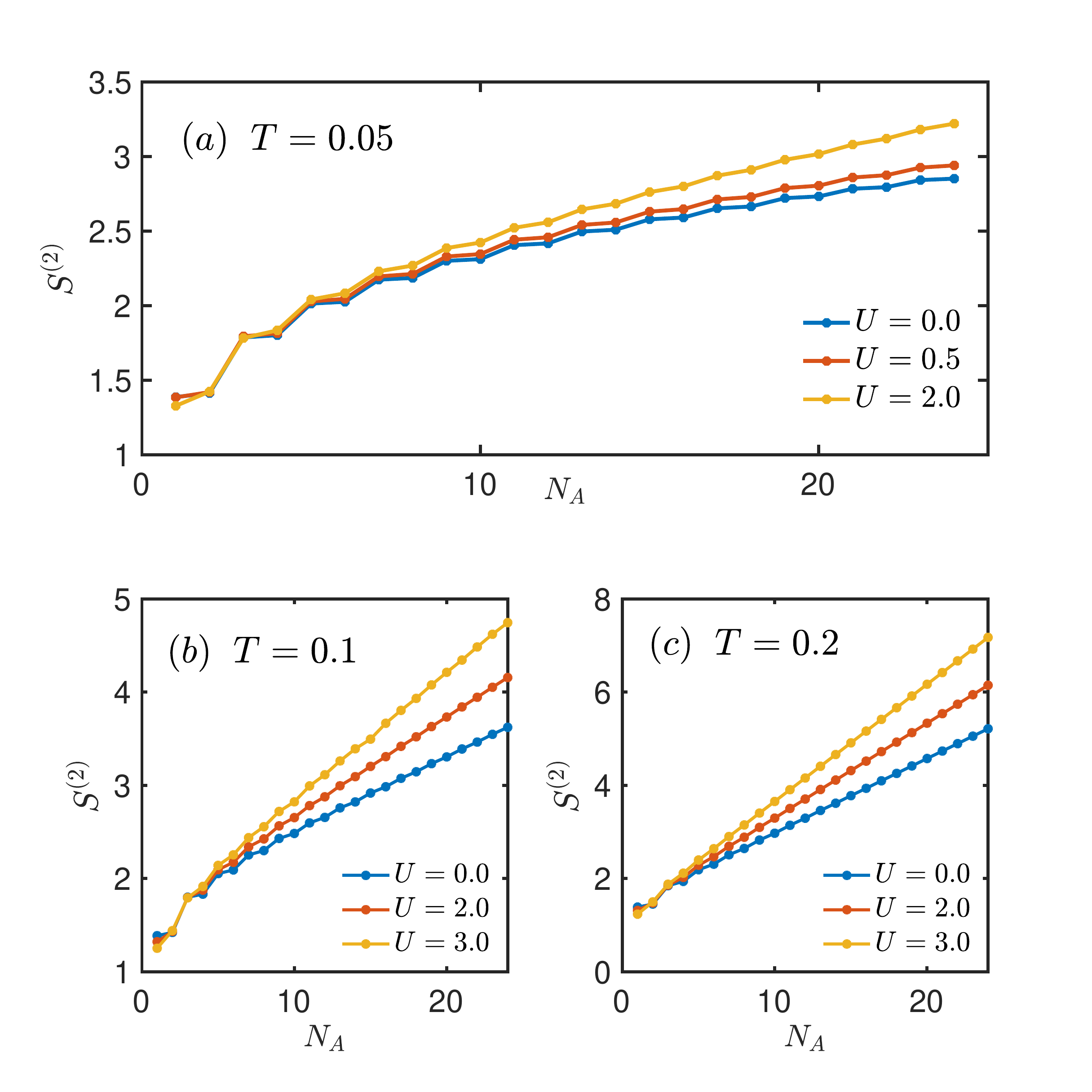}  
\caption{Result for the second R\'{e}nyi entropy $\sr$ in 1d Hubbard model with periodic boundary condition. (a) $\sr$ as a function of subsystem size $N_A$ is shown for $U=0, 0.5, 2$ and total system size $N=50$ at temperature $T=0.05$. $\sr(N_A)$ for $U=0, 2, 3$ and $N=100$ at (b) $T=0.1$, and (c) $T=0.2$. 
The $\sr$ for $U=0.0$ are calculated using the correlation matrix and that for $U\neq 0$ using DMFT.}
\label{mainfig_result_1d}
\end{figure} 

In our DMFT formalism [Sec.\ref{sec:DMFTEntanglement}], the subsystem R\'{e}nyi entropy is obtained from an imaginary-time path integral. Thus we perform the calculations at finite temperature with a finite discretization $\delta \tau$. To obtain the ground-state entanglement, we need to take the $T\to 0$ or $\beta \to\infty$ and $\delta \tau\to 0$ limit. This is not straightforward since at finite temperature $\sr$ contains both thermal and entanglement entropy contributions, and we need to \emph{disentangle} these two contributions as the $T\to 0$ limit is taken. As discussed below, we find that the thermal to entanglement crossover in $\sr$ for the DMFT metallic state can be described by the crossover function known from CFT \cite{Korepin,Cardy,Calabrese_2009,Ryu2006}. 

We show $\sr$ in Fig.\ref{mainfig_result_1d} as a function of subsystem size 
$N_A$ for 1d Hubbard model with periodic boundary condition. In Fig.\ref{mainfig_result_1d}(a), the result for $\sr$ vs. $N_A$ is shown at low temperature $T=0.05$ for the total system size $N=50$ and interaction strengths $U=0, 0.5, 2$. Figs.\ref{mainfig_result_1d}(b, c) show $\sr(N_A)$ at relatively higher temperatures, $T=0.1, 0.2$, for $N=100$ and $U=0, 2, 3$. Here the $U=0$ results are computed using the correlation matrix approach discussed in the preceding section, and $\sr$ for non-zero $U$ is obtained through DMFT.  At higher temperatures, $\sr$ for $N_A\gg 1$ is seen to vary linearly with $N_A$, i.e. a \emph{volume law} scaling. This indicates the dominance of thermal entropy at higher temperatures. An \emph{arc-like} feature emerges at a lower temperatures. This is the hallmark of entanglement contribution to subsystem R\'{e}nyi entropy. Thus, the change of linear to arc-like behavior results from an entropy to entanglement crossover, as we discuss below.

Gapless systems in 1d, like critical bosonic or spin chains, and gapless fermionic chains, exhibit the logarithmic violation of the area-law scaling of entanglement. These systems are usually described by 1+1 D CFT characterized by some central charge $c$ \cite{CardyBook,Cardy}. 
The R\'{e}nyi entropy $S^{(n)}_A$ at $T=0$ for a thermodynamically large system ($N\to\infty$) with one gapless mode is given by the CFT formula for $N_A\gg 1$ \cite{Holzhey1994,Vidal2003,Cardy}
 \beq \label{eq:CFTscaling_1}
 S^{(n)}_A = \frac{1}{2}\bigg(1+\frac{1}{n}\bigg)  \left(\frac{c}{6}\right)\log (N_A) + b'.
 \eeq 
 The logarithmic term above is universal with the central charge $c$, and $b'$ is a subleading non-universal constant originating from high-energy degrees of freedom. 
 For finite $N$, and system with periodic boundary condition \cite{Cardy}, the above formula is modified to 
 \begin{align}  \label{eq:CFTscaling_2}
  S^{(n)}_A = \frac{1}{2}\bigg(1+\frac{1}{n}\bigg) \left(\frac{c}{6}\right)\log\bigg[\frac{N}{\pi}\sin\big(\frac{\pi N_A}{N}\big)\bigg] + b',
 \end{align}
 Similarly, one can obtain the R\'{e}nyi entropy \cite{Korepin,Cardy} at finite temperature and $N\to\infty$ as 
 \begin{align} \label{eq:CFTscaling_3}
    S^{(n)}_A = \frac{1}{2}\big(1+\frac{1}{n}\big) \left(\frac{c}{6}\right)\log\bigg[\frac{v\beta}{\pi }\sinh\bigg(\frac{\pi N_A}{v\beta}\bigg)\bigg] 
   + b   
 \end{align}
where $v$ is a velocity and $b$ is some non-universal constant. For non-interacting spinless fermions, $v=v_\mr{F}$ is the Fermi velocity. In this case, $S_A^{(n)}$ is obtained by adding the contributions of the two gapless chiral modes, i.e., the left ($L$) and right ($R$) movers, at the two Fermi points, each with the central charge $c_L=c_R=c=1$. Eq.\eqref{eq:CFTscaling_3} reduce to Eq.\eqref{eq:CFTscaling_1} for $\beta\to \infty$, i.e., at zero temperature, to give us the ground-state R\'{e}nyi entanglement entropies. For $\beta\to0$, Eq.\eqref{eq:CFTscaling_3} reproduces thermal R\'{e}nyi entropy, $S^{(n)}\simeq (1/2)(1+1/n)(\pi cT/6v)N_A$, e.g. the thermal entropy $S(T)=(\pi cT/6v)N_A$ ($n=1$), scaling linearly with subsystem size. Therefore, the above formula [Eq.\eqref{eq:CFTscaling_3}] can be viewed as a crossover formula\cite{Senthil} from thermal to entanglement entropy. The same low-energy degrees of freedom give rise to the universal part of entanglement and thermal entropy and thus leads to the smooth crossover. The CFT formulas [Eqs.\eqref{eq:CFTscaling_1},\eqref{eq:CFTscaling_2},\eqref{eq:CFTscaling_3}] are also applicable for gapless states of interacting fermions in 1d, i.e., for a Luttinger liquid. In this case, the effect of interaction only enters in the crossover formula [Eq.\eqref{eq:CFTscaling_3}] through the renormalized Fermi velocity $v$, whereas the central charge remains unchanged.

As we discuss in the next section in more detail, for $d>1$, the Fermi liquid state of interacting fermions at $T=0$ also obeys the universal logarithmic scaling \cite{Swingle_PRL} of Eq.\eqref{eq:CFTscaling_1}, which is independent of any Fermi liquid corrections or Landau parameters. The effect of interaction again only appears \cite{Swingle_PRL} in the thermal to entanglement crossover [Eq.\eqref{eq:CFTscaling_3}] through the renormalized $v$. The single-site DMFT [Sec.\ref{sec:DMFTEntanglement}] with the IPT approximation \cite{Kotliar} is designed to give rise to a Fermi liquid metallic state even in 1d. Thus we describe our DMFT results for $S_A^{(2)}(N_A,T)$ (with the spin degeneracy taken into account in it) in the paramagnetic metallic state of 1d Hubbard model using the CFT expression [Eq.\eqref{eq:CFTscaling_3}] with $c$ replaced by $2c$ for the two chiral modes. We note that Eq.\eqref{eq:CFTscaling_3} is valid for thermodynamically large systems ($N\to \infty$). For finite $N$ and $T$, the analytical expression for $S_A^{(n)}(N_A,T,N)$ is not known \cite{Cardy,Ryu2006,Chen2020} to the best of our knowledge. Nevertheless, we use Eq.\eqref{eq:CFTscaling_3} to describe our DMFT data for relatively large systems like $N=50, 100$, as shown in Fig.\ref{mainfig_result_1d}, assuming the $N$ to be large enough for neglecting finite $N$ corrections. Alternatively, we can consider the above crossover function [Eq.\eqref{eq:CFTscaling_3}] as a fitting function. We can fit our $S_A^{(2)}$ as a function of $N_A$ for a fixed $T$ using Eq.\eqref{eq:CFTscaling_3} and three fitting parameters $c$, $v$ and $b$ (Appendix \ref{secsupp:crossover1d}). However, to reduce the number of fitting parameters, we independently extract the ratio $(c/v)$ by fitting the low-temperature specific heat (per site) $c_V$ from equilibrium DMFT calculations [see Appendix \ref{secsupp:cv_eqbDMFT}] with the CFT expression $c_V=(\pi T/3)(c/v)$. 

\begin{figure}[ht!]
    \centering
\includegraphics[width=0.95\linewidth]{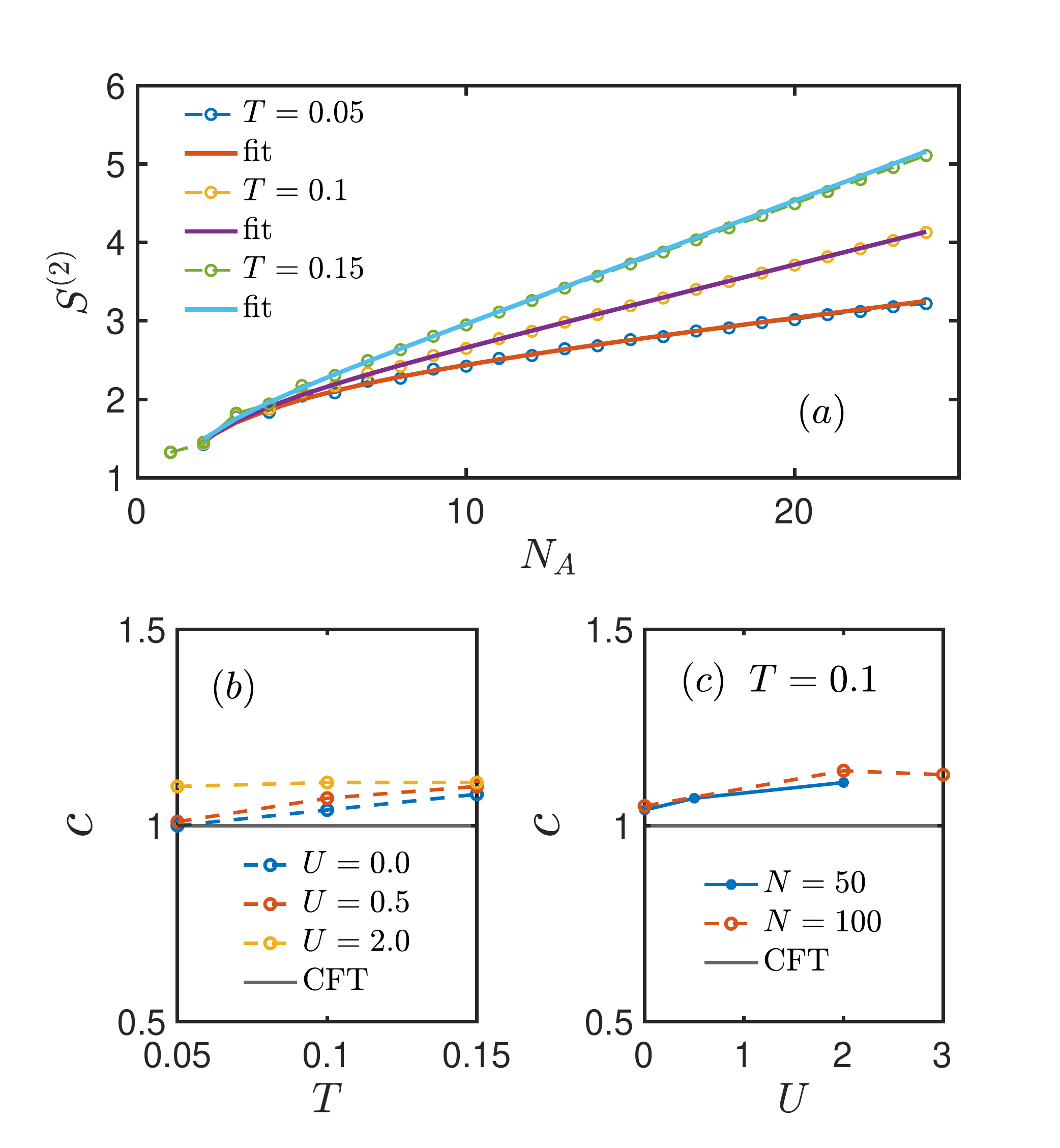}  
\caption{
(a) The second R\'{e}nyi entropy $\sr$ (line + circle) as a function subsystem size $N_A$ for interaction strength $U=2.0$ and system size $N=50$ at  $T=0.05, 0.1, 0.15$ is shown with the fitted CFT crossover function (line) of Eq.\eqref{eq:CFTscaling_3}. (b) The extracted central charge $c$ (dashed line + circle) as a function of $T$ is shown for different interaction strengths $U=0, 0.5, 2$, and compared with the CFT value $c=1$ (solid line). (c) The central charge $c$ as a function of $U$ is shown at $T=0.1$ for two system sizes.}
\label{mainfig_c_sumarry_1d}
\end{figure} 

With the $(c/v)$ ratio fixed this way, we fit Eq.\eqref{eq:CFTscaling_3} to our data with two parameters $c$ and $b$.
As shown in Fig.\ref{mainfig_c_sumarry_1d}(a), $\sr$ follows the crossover function quite well. The extracted central charge $c$ is shown as a function $T$ for fixed $U$, and as a function of $U$ for a fixed $T$ in                                                                                                                                  Figs.\ref{mainfig_c_sumarry_1d}(b) and (c), respectively. The non-universal fitting parameter $b$ and the extracted normalized Fermi velocity $v$ are shown in Appendix \ref{secsupp:cv_eqbDMFT}.
We find that the extracted central charge is close to the free-fermion  value $c=1$. With decreasing temperature the extracted central charge approaches $c=1$ for the non-interacting ($U=0$) and weakly interacting ($U=0.5$) systems [Fig.\ref{mainfig_c_sumarry_1d}(b)], and seems to deviate slightly for relatively stronger interaction $U=2$. However, the deviation might be an artifact of employing the $N\to\infty$ formula [Eq.\eqref{eq:CFTscaling_3}] for finite $N$. In Fig.\ref{mainfig_c_sumarry_1d}(c), we see that $N=50$ and $N=100$ give very similar values of $c\simeq 1$ at $T=0.1$ as a function of $U$, thus assuring convergence with $N$. 
In summary, we conclude that the entanglement properties of DMFT Fermi liquid, captured through $S_A^{(2)}(N_A,T)$, and accessed within the local self-energy approximation match quite well with that of CFT. 

\section{$\sr$ in 2d Hubbard model} \label{sec:2DHubbard}
\begin{figure}[ht!]
\centering
\includegraphics[width=0.80\linewidth]{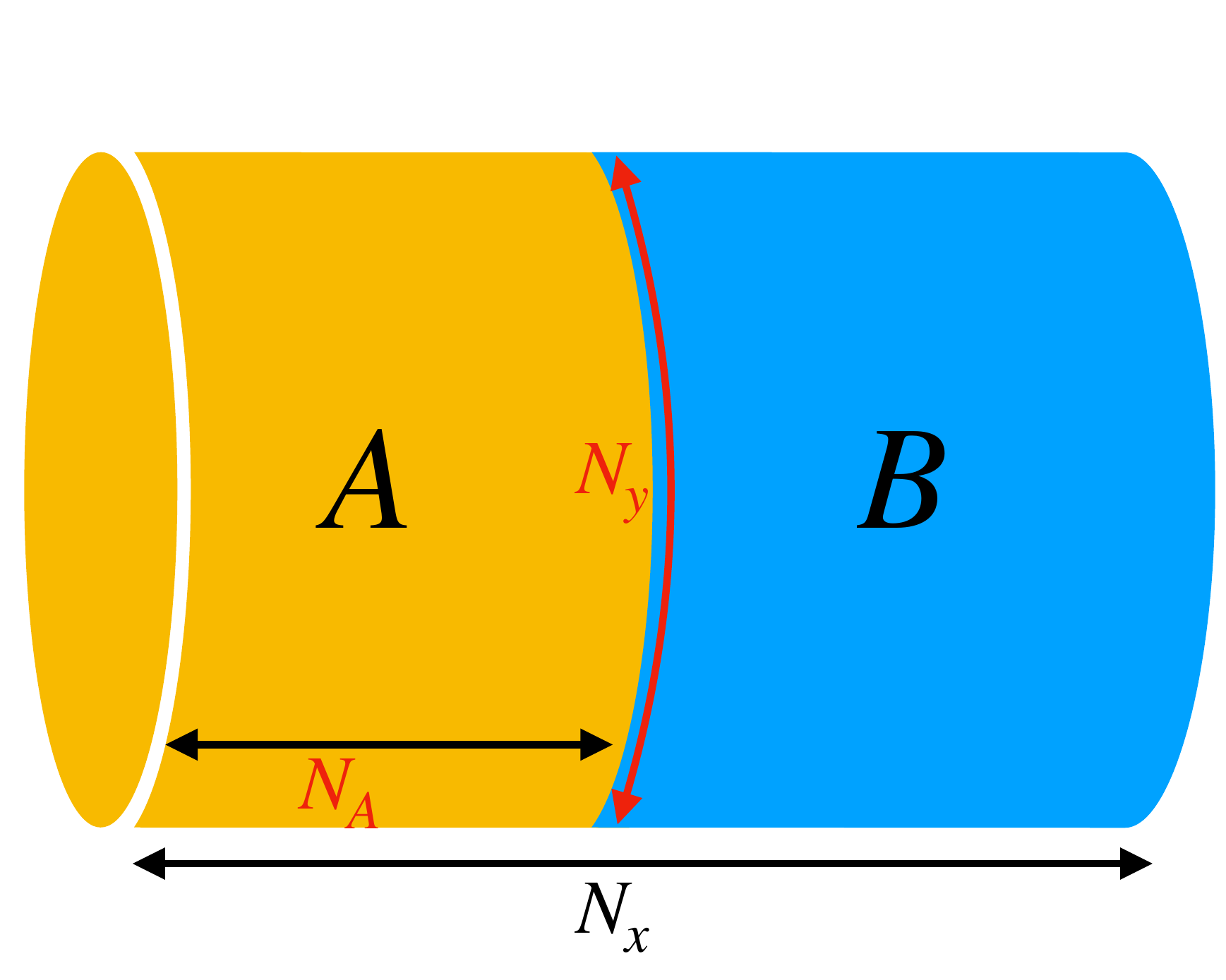}
\caption{The cylindrical subsytem $A$, with $N_A$ sites in $x$ direction and $N_y$ sites in the $y$ direction, used for computing entanglement in 2d Hubbard model, is shown. The periodic boundary condition is applied to the whole system in both $x$ and $y$ directions.
}
\label{supp_fig:subsys2d}
\end{figure}
In this section, we discuss the DMFT results for $\sr$ in 2d Hubbard model for (Fermi liquid) metallic and Mott phases. We consider the system with periodic boundary conditions in both $x$ and $y$ directions, i.e. a torus geometry for the system. We subdivide the system along the $x$ axis, i.e., the entanglement cut is parallel to the $y$ axis like in a  cylindrical subsystem geometry as shown in Fig.\ref{supp_fig:subsys2d}. Due to the entanglement cut we break the translational symmetry in the $x$-direction while retaining the translation symmetry in the $y$ direction, along which periodic boundary condition is applied. As discussed in the Appendix \ref{secsupp:DMFT_2d}, the periodic boundary condition along the $y$-axis allows the wavevector $k_y$ to be a good quantum number, and makes the inversion of the lattice Green's function of Eq.\ref{eq:LatticeGreenEq} easier. In this case, Eq.\ref{eq:LatticeGreenEq} can be decoupled for each $k_y$ mode. 
We discuss below our DMFT results for $\sr$ and its dependence on the subsystem size $N_A$, temperature, and interaction strength. 

Most systems in higher dimension ($d>1$) follow boundary law scaling of entanglement. Even the higher dimensional CFT gives strict boundary law scaling. However, there are several important exceptions \cite{Swingle_PRL,Swingle2012,Swingle2012a,Senthil} with logarithmic violation of the boundary law, e.g., free fermions, Fermi liquids, Weyl fermions in a magnetic field, non-Fermi liquids with critical Fermi surface, and Bose metals. Here we focus on Fermi liquid metallic state as captured within DMFT. The underlying reason behind the violation of the area law is that these systems with Fermi surface can be effectively described as a collection of patches on the Fermi surface \cite{Swingle_PRL,multibosonization}. Each of these Fermi surface patches acts as a one-dimensional gapless chiral mode described by 1+1 D CFT. These modes are chiral as they can only propagate with Fermi velocity radially outward to Fermi surface at very low temperatures. Then, the scaling of entanglement entropy with $N_A$ is simply the one-dimensional logarithmic scaling multiplied by the number of gapless 1+1 D CFT modes \cite{Swingle_PRL}. The counting of the number of these mode depends on both the geometry of the Fermi surface and real space boundary \cite{Swingle_PRL}. 

As discussed in the preceding section, for one dimension, we have both right and left movers mode with central charge $c_L=c_R=c=1$ and the scaling of R\'{e}nyi entanglement entropy is given by Eq.\ref{eq:CFTscaling_1}. For the chiral mode in $d>1$, we have either $c_L=0,~c_R=c$ or $c_R=0,~c_L=c$, and hence the contribution (per spin component) of each chiral mode to the R\'{e}nyi entropy at $T=0$ is still given by Eq.\eqref{eq:CFTscaling_1}. 
The counting of the mode is obtained from the Widom formula \cite{Gioev2006,Calabrese2009a,Swingle_PRL,Swingle2012,Swingle2012a,Leschke2014}, originally developed in the context of signal processing \cite{Widom1982},
\begin{align}\label{eq:Nmodes}
N_{\rm modes} = \frac{1}{(2\pi)^{d-1}}\frac{1}{2}\int_{\partial A_x}\int_{\partial A_k} dA_x dA_k |\hat{\boldsymbol{n}}_x\cdot \hat{\boldsymbol{n}}_k| .
\end{align}
The integrals are over the real-space boundary $\partial A_x$ of the subsystem and the Fermi surface $\partial A_k$. $\hat{\boldsymbol{n}}_x$ and $\hat{\boldsymbol{n}}_k$ are the unit normals to the real-space boundary and the Fermi surface, respectively. Here, the \emph{flux factor} $|\hat{\boldsymbol{n}}_x\cdot \hat{\boldsymbol{n}}_k|$ counts the fraction of modes perpendicular to real-space boundary coming from a Fermi surface patch at $\boldsymbol{k}$. The Widom formula has been verified numerically \cite{Li2006} for free fermions in $d>1$. For Fermi liquids, where only forward scattering is relevant, the Widom formula is expected to remain valid \cite{Swingle_PRL,multibosonization} with the same $c$ in Eq.\eqref{eq:CFTscaling_1} modulo possible modification of the Fermi surface geometry due to interaction if any. Going beyond Fermi liquids, the Widom formula may get violated \cite{Haldane,Si} or modified \cite{Senthil}, e.g., as in the case of gapless states of composite fermions in the fractional quantum Hall regime, quantum spin liquids, and non-Fermi liquids.


For the square lattice Hubbard model [Eq.\eqref{eq:HubbardModel}] that we consider here, the non-interacting dispersion is $\varepsilon_k=-2t(\cos k_x + \cos k_y)$. Thus we can compute the $N_{\rm modes}$ for the cylindrical subsystem ($N_A\times N_y$) as discussed in Appendix \ref{secsupp:widomformula}. In this case, $N_{\rm modes}$ is given by $2N_y$, where $N_y$ is the number of sites in the $y$ direction along the entanglement cut. Therefore, taking the spin degeneracy into account, we expect
\begin{align}
    \sr/N_y =  \frac{c}{2}\log (N_A) + b'
\end{align}
at $T=0$. 
Moreover, like in the 1d case [Eq.\eqref{eq:CFTscaling_3}], we expect entropy to entanglement crossover at finite temperature for a thermodynamically large system to be given by
\begin{align}\label{eq:2dcrossover_finalform}
      \sr/N_y = \frac{c}{2}\log\bigg[\frac{v\beta}{\pi}\sinh \big(\frac{\pi N_A }{v\beta}\big) \bigg] + b  
\end{align}
where $b$ is again another non-universal constant. Hence the R\'{e}nyi entropy per unit length along $y$-direction i.e $ \sr/N_y$ in this cylindrical subsystem geometry almost have same form as the 1d crossover formula [Eq.\eqref{eq:CFTscaling_3}]. As in the 1d case, we do not have any crossover formula that interpolates between entanglement and thermal entropy for finite $N$ and finite temperature. For the correlated metallic state obtained in our DMFT calculations for the 2d Hubbard model, we verify the crossover formula [Eq.\eqref{eq:2dcrossover_finalform}].
Again the effect of interaction only enters in the crossover formula via the velocity $v$ for a Fermi liquid.
 We note that a universal entropy-entanglement crossover formula may be valid more generally, even beyond Fermi liquids. Similar crossover formulae, constrained by the temperature dependence of thermal entropy, have been proposed \cite{Senthil} to hold even for gapless fermionic systems not described by Fermi liquid theory or devoid of quasiparticles, e.g., non-Fermi liquids with critical Fermi surfaces. 

\begin{figure}[ht!]
\centering
\includegraphics[width=0.95\linewidth]{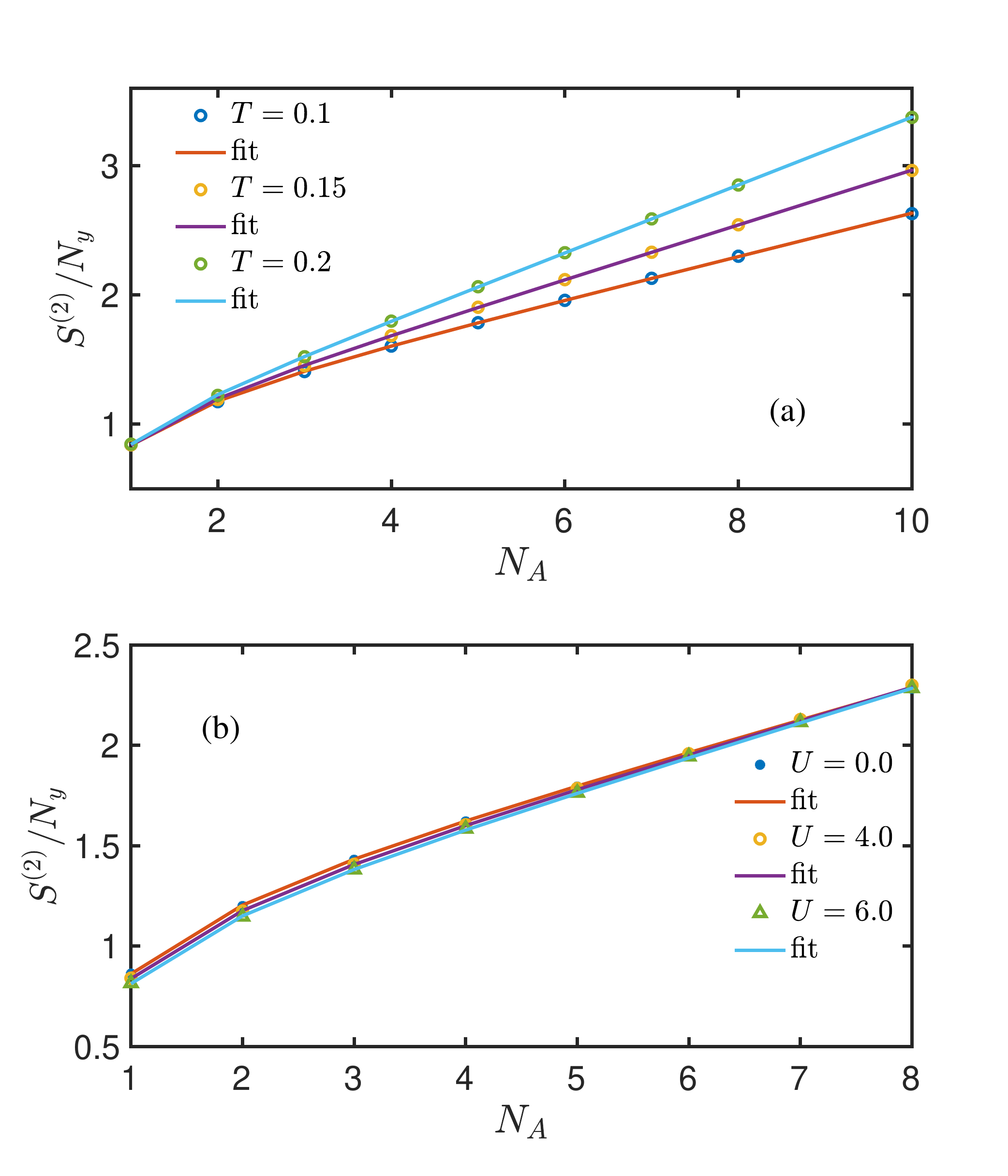}  
\caption{
(a) $\sr/N_y$ for the metallic state of 2d Hubbard model is shown for a $N=20\times 20$ lattice as a function of the subsystem length along $x$-axis $N_A$ for temperatures $T=0.1, 0.15, 0.2$ at fixed interaction strength $U=4$. The results (circle) fit well with the crossover function (line) of Eq.\eqref{eq:2dcrossover_finalform}. 
(b) $\sr/N_y$ (fill circle, open circle, triangle) as a function of $N_A$ is shown for different interaction strengths $U=0, 4, 6$ at fixed $T=0.1$.  The corresponding fits (line) to the crossover function are also shown.}
\label{mainfig_2dresult}
\end{figure} 
 
We compute the second R\'{e}nyi entropy $\sr$ as a function of subsystem size $N_A N_y$ for a total system size  $N_x N_y$. Here $N_A$ is the number of lattice sites in the subsystem $A$ in the $x$ direction, and $N_x$ and $N_y$ are the total number of sites in the $x$ and $y$ directions, respectively. We vary the subsystem size by varying $N_A$ while keeping $N_y$ fixed. In Fig.\ref{mainfig_2dresult}(a), $\sr/N_y$, computed from DMFT, is shown as a function of $N_A$ at low temperatures, $T=0.1, 0.15, 0.2$, for $U=4$ and system size $N=20\times 20$. As shown in the figure, we fit the DMFT results with the crossover formula of Eq.\eqref{eq:2dcrossover_finalform}. We find the formula to fit very well to our result as shown in the [Fig.\eqref{mainfig_2dresult}(a)].  Similarly, in Fig.\eqref{mainfig_2dresult}(b), the $\sr/N_y$ is shown for different interaction strengths $U=0, 4, 6$ at fixed $T=0.1$ with the corresponding fits to the crossover formula [Eq.\eqref{eq:2dcrossover_finalform}] for a $20\times 20$ system. 

\begin{figure}[ht!]
\centering
\includegraphics[width=0.95\linewidth]{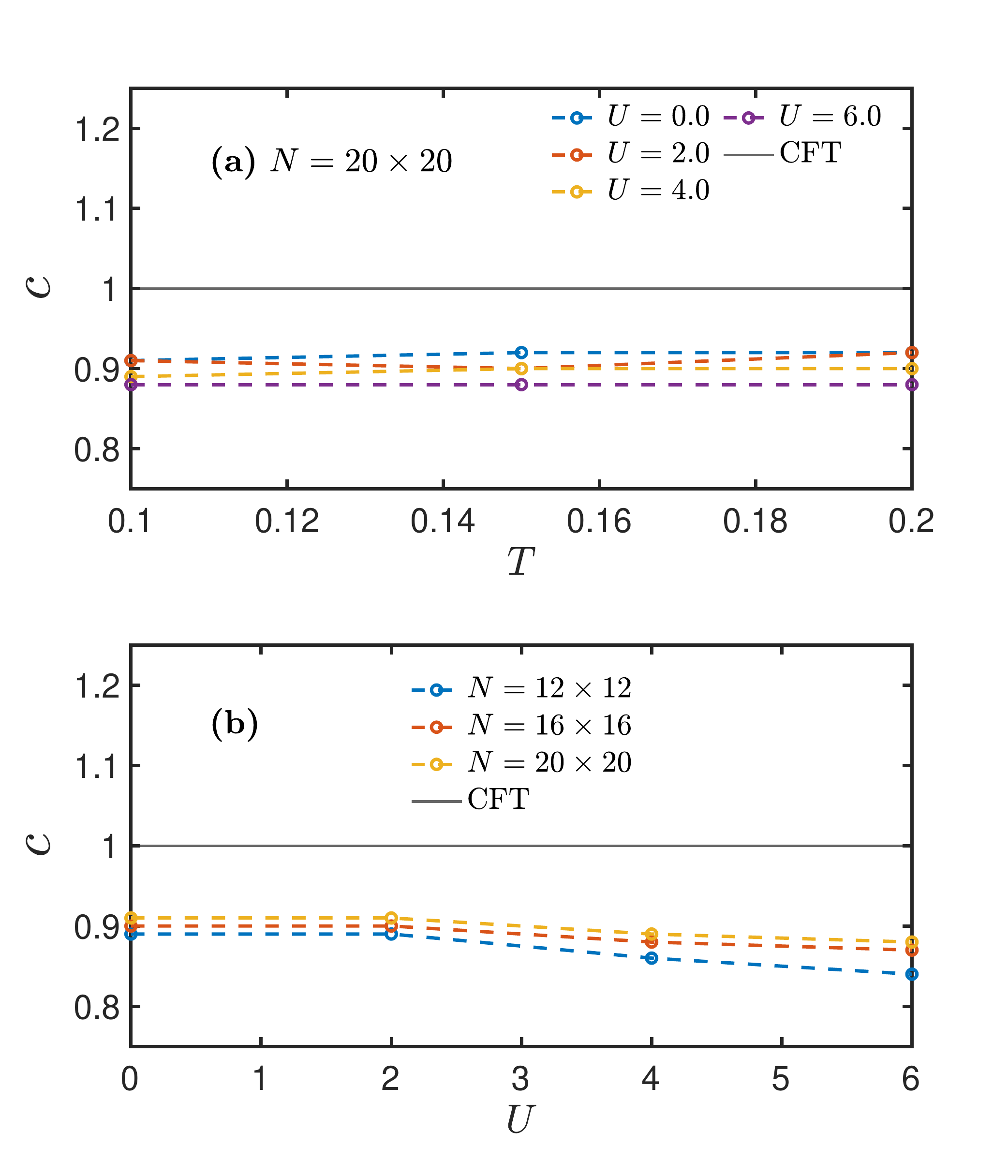}  
\caption{ (a) The extracted central charge $c$ as a function of $T$ is shown for different interactions $U=0, 2, 4, 6$ and system size $N=20\times 20$, and compared with the CFT value $c=1$.  (b) The variation of the $c$ as a function of $U$ is shown at fixed $T=0.1$ for different system sizes $N$, as indicated in the legends.}
\label{mainfig_2d_c_summary}
\end{figure}

The $c$ extracted from the above fittings is shown in Fig.\ref{mainfig_2d_c_summary}. More details are given in Appendix \ref{secsupp:widomformula}. Fig.\ref{mainfig_2d_c_summary}(a) shows the extracted $c$ as a function of $T$ for $U=0,2, 4, 6$, and compares with the expected CFT value $c=1$. We see that the calculated $c\simeq 0.9<1$, and $c$ does not vary much with $T$. We find that $c\simeq 0.9<1$ even for $U=0$, i.e., $c$ deviates from 1 by more or less the same amount even for the non-interacting case for the system sizes accessed. Thus, presumably, the deviation from the CFT value stems from the application of the crossover formula [Eq.\eqref{eq:2dcrossover_finalform}] for the thermodynamic limit to the finite systems. In Fig.\ref{mainfig_2d_c_summary}(b), where we plot $c$ as a function of $U$ for different $N$ and $T=0.1$, we observe that $c$ decreases slightly for larger $U$. $c$ tends to increase very slowly with increasing $N$, implying that $c$ might approach the expected value of $1$ for larger systems. However, in the absence of an analytical crossover function for finite $N$ and $T$, and for the accessible system sizes in our calculations, it is hard to extrapolate $c$ to the $N\to \infty$. Nevertheless, we can conclude that modulo finite-size effects, our DMFT results for $\sr (N_A,T,N)$ for the metallic state of the 2d Hubbard model at half filling are consistent with the Widom formula and the expectations from CFT \cite{Swingle_PRL}.

\begin{figure}[ht!]
    \centering
\includegraphics[width=0.90\linewidth]{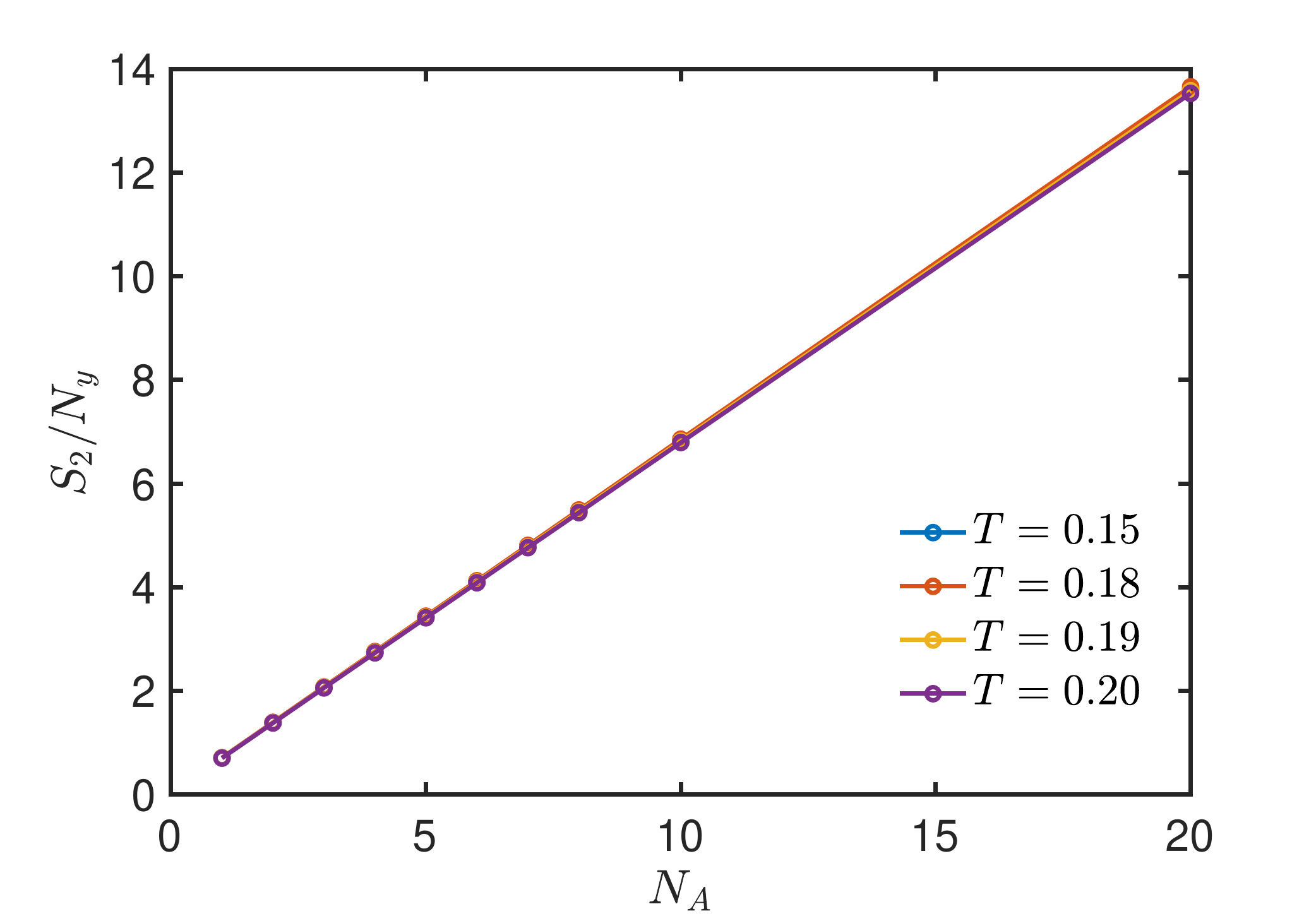}
\caption{The $\sr/N_y$ as a function of $N_A$ is shown for several $T$'s at fixed $U=14$. This is computed from system size $20\times 20$. }
\label{supp_fig:S2Mottphase}
\end{figure}

We also compute the R\'{e}nyi entropy deep inside the insulating phase at low temperature, as shown in Fig.\ref{supp_fig:S2Mottphase} for $U=14$. We find that $\sr/N_y$ is linear in $N_A$ with a slope $\gtrsim \ln 2$, with a very weak dependence on $T$. This is expected for $T\ll U$, due to non-zero $T\to 0$ residual entropy\cite{Georges}, arising from spin degeneracy of the paramagnetic Mott insulating solution in the single-site DMFT, .  




\section{Mutual information across Mott transition in 2d Hubbard model} \label{sec:MutualInfo}

\begin{figure}[ht!]
\centering
\includegraphics[width=0.9\linewidth]{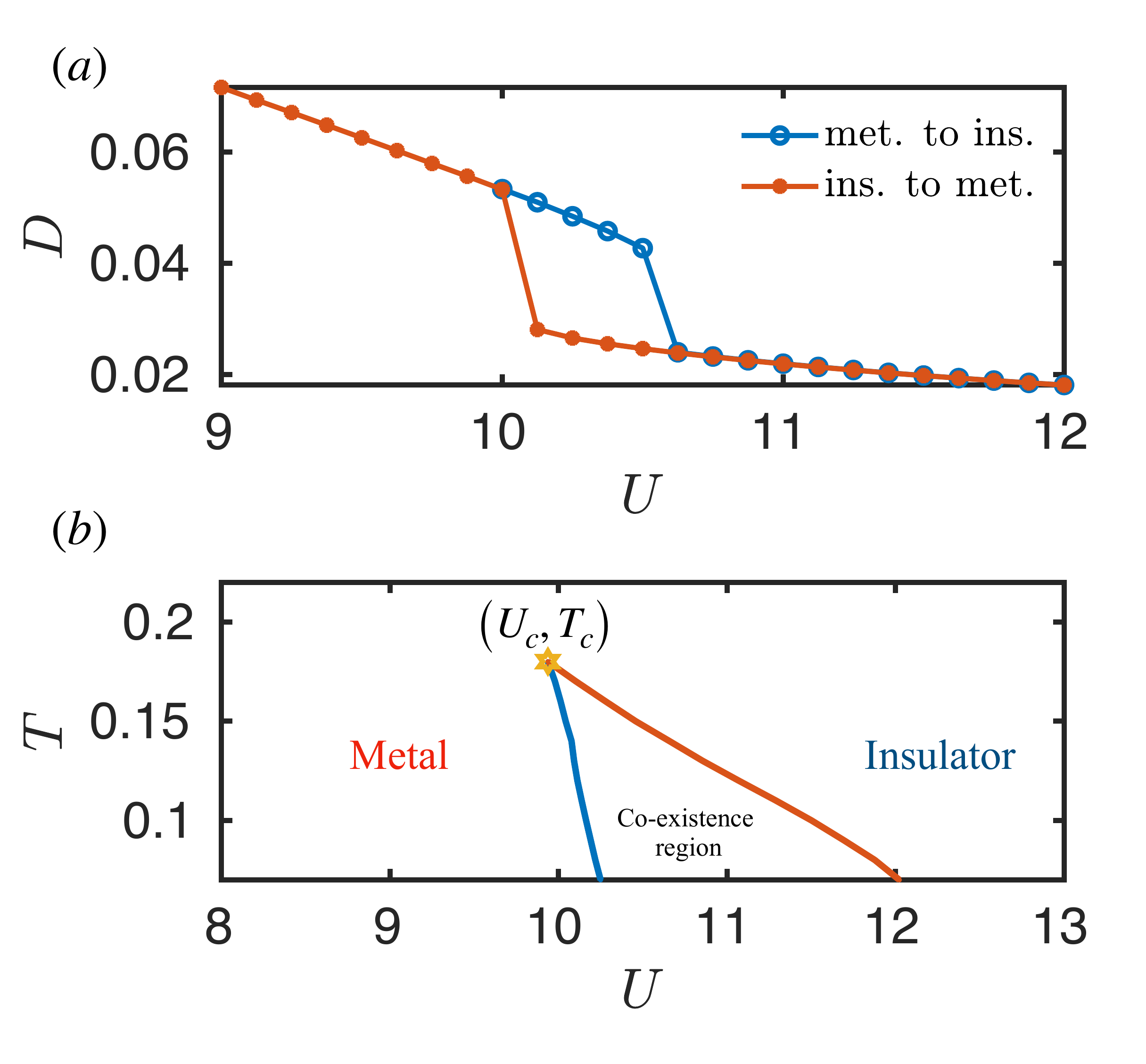}
\caption{(a) Double occupancy $D$, computed within single-site DMFT, as a function of $U$ is shown at $T=0.14$ for 2d Hubbard model. The legend `met. to ins.' implies increasing $U$ from the metal to insulator and `ins. to met.' decreasing $U$ the other way round. (b) The phase diagram for 2d half-filled Hubbard model constructed based on the double occupancy $D$ is shown. $(U_c, T_c)$ represents Mott critical point where the first-order line, or the coexistence region, ends. In the co-existence region, both metal and insulator solutions exist.}
\label{mainfig_phasediag}
\end{figure} 

Here we discuss the second R\'{e}nyi mutual information as an entanglement and information theoretic measure at finite temperature in the temperature vs. interaction ($T-U$) phase diagram  of the Hubbard model. The second R\'{e}nyi mutual information, 
\begin{align}
I(A, B)&=S^{(2)}_A+S^{(2)}_B-S^{(2)}_{A\cup B} \nonumber \\
&=-(\ln Z_A^{(2)}+\ln Z_B^{(2)}-\ln Z_{A\cup B}^{(2)}+2 \ln Z) \label{eq:MuInfo}
\end{align}
is obtained from the combination of the R\'{e}nyi entropies of a subsystem $A$, its complement $B$, and the whole system $A\cup B$. While the entanglement entropy can characterize pure states, e.g., ground state and quantum phase transitions between ground states at $T=0$, the mutual information is a better information theoretic measure for finite-temperature phases and phase transitions \cite{Melko2010,Singh2011,Iaconis2013,Stephan2014}. The mutual information is dominated by entanglement contribution when classical correlations are short-ranged, e.g., at low temperature away from a finite-temperature critical point \cite{Wolf2008}. Moreover, different parts of mutual information can exhibit critical properties \cite{Melko2010,Singh2011,Iaconis2013,Stephan2014} at temperatures related to the finite-temperature critical point, e.g., at critical temperature $T_c$, and at $2T_c$ due to critical behaviours of $Z_A^{(2)},~Z_B^{(2)}$ in Eq.\eqref{eq:MuInfo} from the edges and corners of the subsystem $A,~B$.

 For a pure-state density matrix, $I(A,B)=2\sr$. The $\sr$ for thermal density matrix at finite temperature contains both entanglement and thermal entropy contributions. However the (R\'{e}nyi) mutual information, by construction, naturally excludes the volume-law thermal entropy of the subsystem and its complement. Thus mutual information follows in general an area-law scaling with subsystem size and captures both quantum (entanglement) and classical correlations between the subsystems. The study of Mott transition through the lens of mutual information is less explored in literature. In refs.\onlinecite{Tremblay,Walsh2019,Walsh2020}, the authors have studied the Mott transition in the 2d Hubbard model using equilibrium CDMFT through the mutual information of a single site and the rest of the system. They detect first-order phase transitions and supercritical regime for $T>T_c$. The calculation of the singe-site mutual information only requires the knowledge of occupation and double occupancy, that can be computed within equilibrium DMFT. The subsystem size scaling of mutual information cannot be captured within such equilibrium DMFT. Within our new path integral approach we can easily study the subsystem size scaling of mutual information across Mott transition, as we discuss below. 

\begin{figure}[ht!]
\centering
\includegraphics[width=0.9\linewidth]{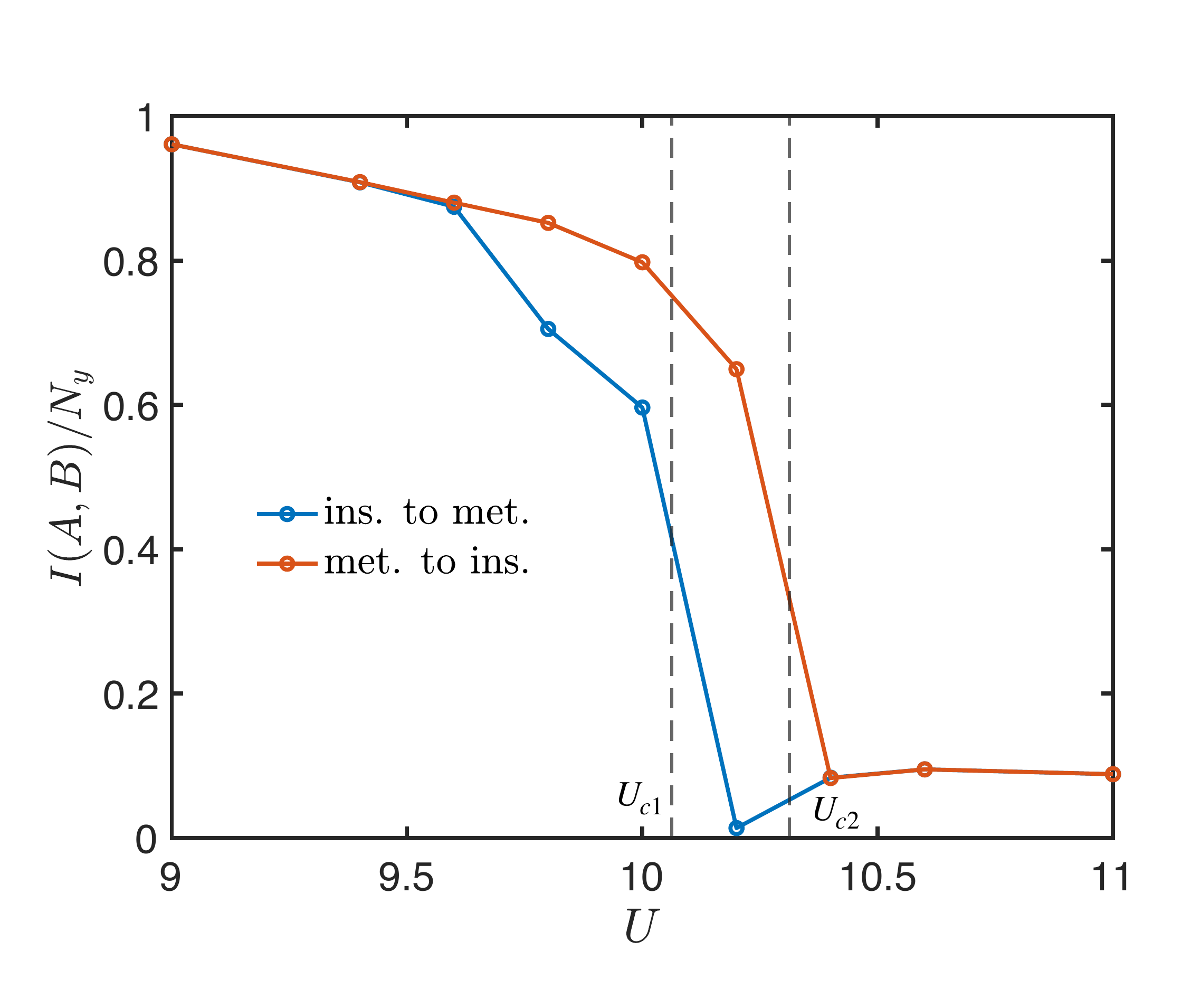}
\caption{The bipartite mutual information for $N_A=N_B$ per site of the subsystem along $y$ axis, i.e., $I(A, B)/N_y$ as a function of interaction $U$ at $T=0.16<T_c$ is shown for a $N=16\times 16$ lattice. `Met. to ins' refers to sweeping $U$ from the metal to insulator phase and `ins. to met' for the other way round.
}
\label{mainfig_hysteresis}
\end{figure} 
 
To characterize the metal, insulator and the metal-insulator phase transition at finite temperature, we first briefly discuss the $T-U$ phase diagram of the half-filled 2d Hubbard model within large connectivity Bethe lattice approximation in DMFT\cite{Georges}. We draw the $T-U$ phase diagram by monitoring the equal time correlation function $D=\langle n_{\uparrow}n_{\downarrow} \rangle$ corresponding to the double occupancy  of a site within equilibrium DMFT (Appendix \ref{secsupp:cv_eqbDMFT}) as a function of $U$ for different temperatures. A representative plot for double occupancy $D$ vs. $U$ at $T=0.14$ is shown in Fig.\ref{mainfig_phasediag}(a). The hysteresis behaviour is due to the coexistence of both metal and insulator solution across the first-order Mott metal-to-insulator transition. The area of the hysteresis loop shrinks to zero as the first order transition ends at finite temperature Mott critical point ($U_c, T_c$). By monitoring $D(U)$, we obtain the $T-U$ phase diagram, as shown in Fig.\ref{mainfig_phasediag}(b). The critical point is at $(U_c, T_c)\approx (9.9, 0.18)$. 


In Fig.\ref{mainfig_hysteresis}, the mutual information per site along the $y$ durection for the equal bi-partition $N_A=N_B$, i.e., $I(A,B)/N_y$ is shown at $T=0.16<T_c$ for system size $N=16\times 16$. As mentioned earlier, for the first-order Mott transitions, two co-existing solutions, metal and insulator, appear in the phase diagram as indicated in Fig.\ref{mainfig_phasediag}(b). In equilibrium DMFT, for $T<T_c$, starting from the insulating solution at large $U$ and on decreasing the interaction slowly, a sudden jump to the metallic solution occurs at $U_{c1}(T)$, i.e., at the limit of metastability of the insulating phase, as usual in a first-order transition. Similarly, sweeping $U$ from the metallic side leads to a jump to the insulating solution at $U_{c2}(T)$. For $T=0.16$, $U_{c1}$ and $U_{c2}$ computed from double occupancy within the equilibrium calculation are shown in Fig.\ref{mainfig_hysteresis} as dotted vertical lines. The DMFT for R\'{e}nyi entropy also leads to similar hysteresis behaviour in the mutual information, as shown in Fig.\ref{mainfig_hysteresis}, where the calculation of $I(A,B)$ vs. $U$ is done in steps of $\delta U = 0.2$. Following the behaviour of double occupancy, the bipartite mutual 
information $I(A,B)/N_y$ also jumps across $U_{c1}, U_{c2}$. Thus, the  mutual information between two extended subsystems can detect the first order nature of phase transitions, like the single-site mutual information \cite{Tremblay,Walsh2019,Walsh2020}. Finite but weak correlations, indicated by the non-zero mutual information, persist even in the insulating phase for $U\gtrsim U_{c2}$, as can be seen in Fig.\ref{mainfig_hysteresis}. We expect these correlations to approach zero for large interaction strengths $U\gg W$, where $W$ is the non-interacting bandwidth.

The calculation of mutual information near the critical point $(U_c,T_c)$ through the extrapolation $\delta\tau \to 0$ becomes challenging due to multiple solutions as well as very close numerical values of  $S^{(2)}(\delta\tau)$ for different $\delta\tau$'s. For this reason, we present the data for a fixed $\delta\tau = 0.029$, without any extrapolation, in this section.


\begin{figure}[ht!]
\centering
\includegraphics[width=0.95\linewidth]{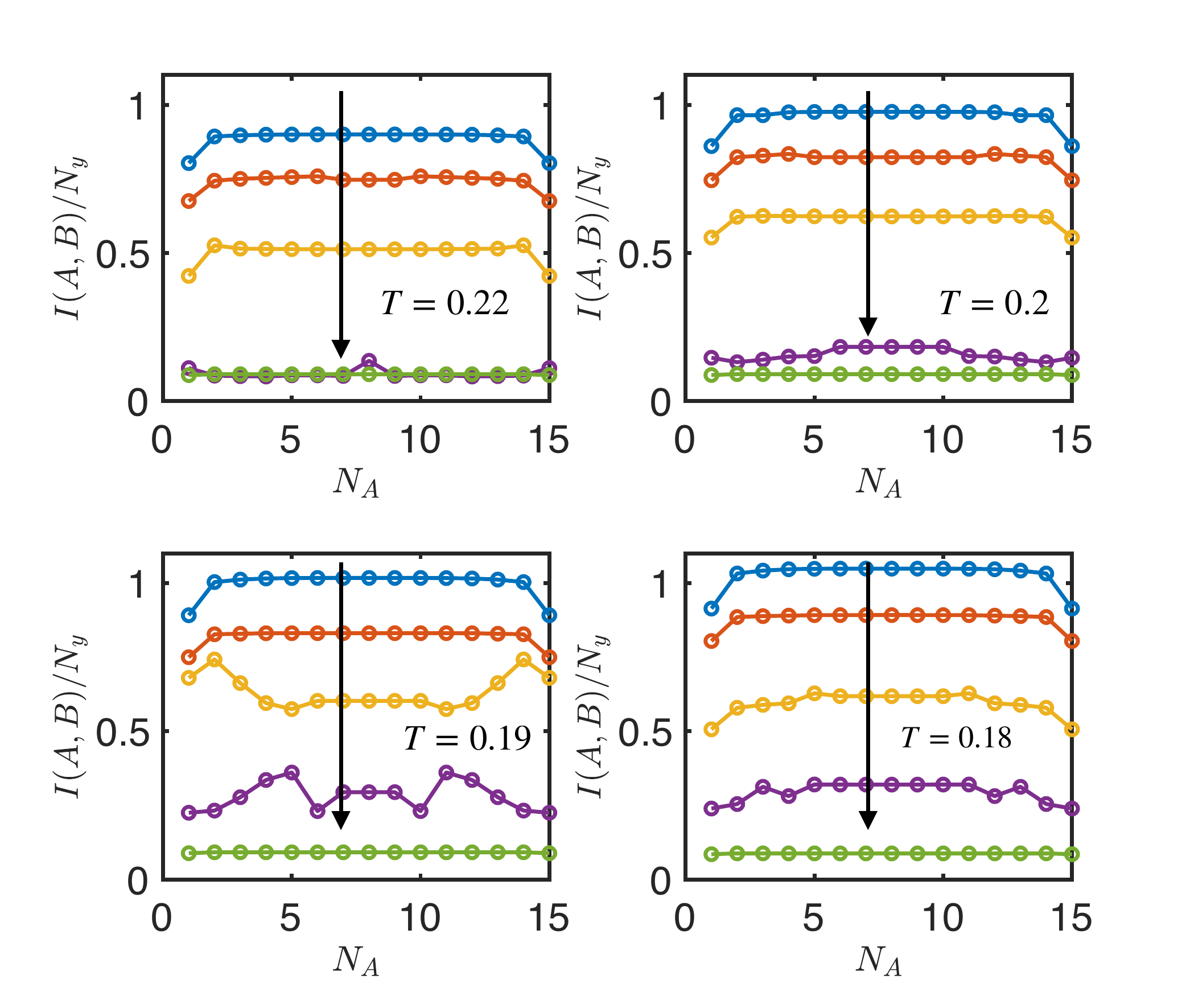}
\caption{The mutual information $I(A,B)/N_y$ for $N=16\times 16$  as a function of the length of subsystem $N_A$ along $x$-direction is shown here across the Mott metal-insulator transition for different interaction strength $U$ ($U=8.0, 9.0, 9.6, 10.0, 11.0$) and temperatures. $U$ gradually increases from top to bottom, as indicated by the arrow, in each plot for a fixed temperature.}
\label{mainfig_subsysscaling}
\end{figure} 

The subsystem size scaling of mutual information is expected to be of the form\cite{Melko2010,Singh2011} $I(A,B)=I(N_A, N_y, \beta) \simeq a(\beta, N_A) N_y + d(\beta, N_A) + \mathcal{O}(N^{-1}_y)$, with coefficients $a$ and $b$ weakly dependent on $N_A$. 
In our subsystem geometry, the dominant contribution to $I(A,B)$ comes from the interface of the two subsystem $A$ and $B$, and leads to the leading area law ($\propto N_y$) for the mutual information with the coefficient $a(N_A, \beta)$. The latter is expected to approach a constant value with subsystem size $N_A$ for sufficiently large $N_A$.
The other term $d(N_A, \beta)$ can appear due to the corner contribution or, for a finite system, from the degeneracy of thermodynamic state, arsing from symmetry breaking\cite{Melko2010,Singh2011} or configurational entropy. In our subsystem geometry [Fig.\ref{supp_fig:subsys2d}], the corner contribution is absent. The finite-temperature Mott transition is similar to a liquid-gas transition \cite{Kotliar2000}. Thus, for the second R\'{e}nyi mutual information, constant term $d(\beta,N_A)$ can appear between $T_c<T<2T_c$  and $T<T_c$ from effective Ising-like symmetry breaking in different parts of $I(A,B)$ along the first-order transition line in the $T-U$ plane. 
We show $I(A,B)/N_y$ for $N=16\times 16$  lattice as a function of $N_A$ in Fig.\ref{mainfig_subsysscaling} for different $U$s across Mott transition at several temperatures near Mott critical point. The arrows in the plots indicate increasing values of interaction over the range $U=8-11$. 
In Fig.\ref{mainfig_subsysscaling}, we see that $I(A,B)/N_y$ becomes more or less independent of $N_A$ for $N_A\sim N/2$, except near the critical point $(U_c,T_c)$, where more complex dependence on $N_A$ is seen. 
\begin{figure}[ht!]
    \centering
\includegraphics[width=0.90\linewidth]{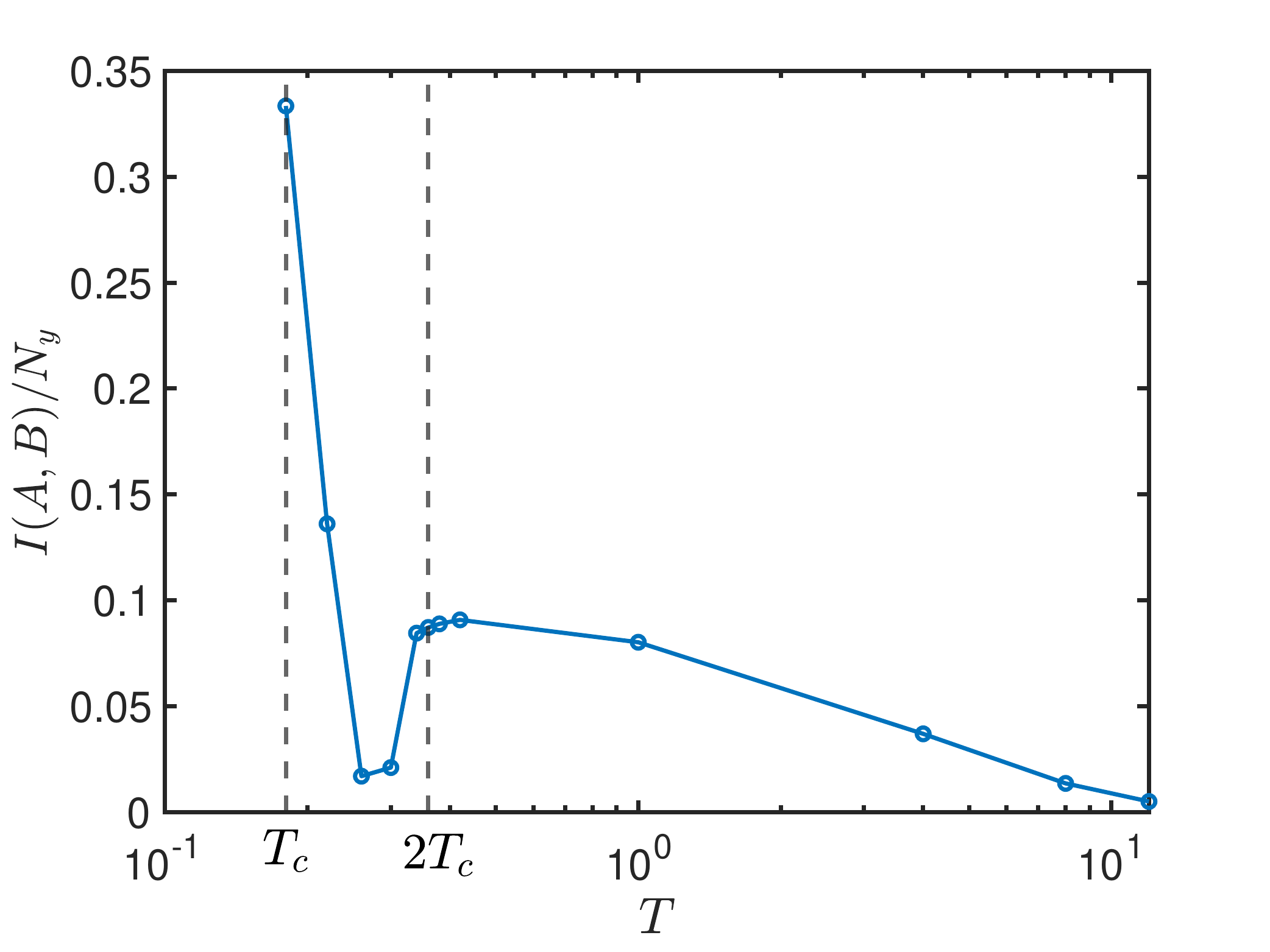}
\caption{The non-monotonic temperature dependence of $I(A,B)/N_y$ is shown for $U=10.0\approx U_c$. This is computed from system size $16\times 16$. Here $T$ is plotted in the logarithmic scale.}
\label{main_fig:IABvsT}
\end{figure}

Finally, in the Fig.\ref{main_fig:IABvsT}, we show the mutual information $I(A, B)/N_y$ for the bi-partition $N_A=N_B$ as a function of temperature (in logarithmic scale) near $U\simeq U_c=10.0$. The $I(A, B)$ shows a non-monotonic behaviour with $T$. In particular the mutual information appears to dip between $\sim T_c$ and $\sim 2T_c$. A substantial $I(A,B)$, indicating correlations, persists even far above $T_c$ till $T\lesssim W$ in the supercritical regime \cite{Tremblay,Walsh2019,Walsh2020}. In future, it will be interesting to study the system-size scaling of $I(A,B)$ with $N_y$ or $N$ to understand the nature of correlations in the supercritical regime contributing to the mutual information of an extended subsystem.

\section{Conclusions and discussion} \label{sec:Conclusion}
To summarize, we have developed a DMFT formalism and its numerical implementation for computing R\'{e}nyi entanglement entropy and mutual information in strongly correlated electronic systems described by the Hubbard model. We show that the scaling of the R\'{e}nyi entropy with subsystem size for an extended subsystem can be used to characterize correlated Fermi liquid metallic state in the half-filled Hubbard model. In particular, we show that the subsystem-size scaling of R\'{e}nyi entropy follows the entropy to entanglement crossover formula expected from CFT and related arguments even in the presence of strong electronic correlations captured by local self-energy approximation within the single-site DMFT. We also show how the first-order Mott transition and Mott critical point are manifested in the temperature, interaction and subsystem-size dependence of R\'{e}nyi mutual information.

Here, as a first attempt to implement the entanglement path integral formalism of ref.\onlinecite{Haldar} within DMFT, we use an approximate impurity solver, namely the IPT \cite{Georges}. An immediate extension of our work would be to employ the CTQMC impurity solver\cite{Gull} in the entanglement DMFT framework. Our entanglement path integral formalism is naturally suited for such a purpose and only requires the incorporation of the local kick self-energy [Eq.\eqref{eq:DMFTeq1}] in the impurity action for the CTQMC solver\cite{Gull}. Another interesting, albeit more challenging, future direction would be to capture short-range correlations via cluster extension\cite{Kotliar2001,Kotliar} of the DMFT formalism. On a different note, it will be interesting to explore the connections between real-space and momentum-space\cite{Flynn2023} entanglement in a Fermi liquid.

In this work, we have discussed the implementation of the entanglement path integral for thermal density matrix suitable to approach entanglement of many-body ground states of the Hubbard model in the $T\to 0$ limit. However, like the usual non-equilibrium DMFT\cite{RMPNonEq}, the entanglement path integral and the DMFT can be extended to non-equilibrium situations via Schwinger-Keldysh formalism, as discussed in ref.\onlinecite{Haldar,Moitra}. Moreover, the non-equilibrium formulation can be further generalized to incorporate non-unitary dynamics, e.g., the dynamics and the steady states of Hubbard model under repeated projective or weak measurements, to study entanglement transitions similar to that seen in random quantum circuits\cite{Fisher2022}. In recent years, DMFT, with its integration with other first-principle electronic structure methods \cite{Kotliar}, has become one the most practical approaches to describe realistic strongly correlated systems. Thus, our DMFT formulations, along with its possible extensions discussed above, might lead to a viable route to computing entanglement properties of strongly correlated materials in future.


\begin{acknowledgements}
SB acknowledges support from SERB (ECR/2018/001742), DST, India and QuST, DST, India. 
\end{acknowledgements}

\appendix
\renewcommand{\thefigure}{A\arabic{figure}}
\setcounter{figure}{0}

\section{Numerical solution of the DMFT equations} \label{secsupp:DMFT_A}

The DMFT equations [Eqs.(\ref{eq:DMFTeq1}, \ref{eq:DMFTeq2}, \ref{eq:DMFTeq3}, \ref{eq:LatticeGreenEq}, \ref{eq:hybridization_eq1}, \ref{eq:BetheApprox})] for computing $\sr$ are numerically much more challenging to solve compared to usual equilibrium DMFT equations \cite{Georges} since the Green's function $G_{i\sigma\alpha,j\sigma\beta}(\tau,\tau')$ [Eq.\eqref{eq:RenyiGreenFn}] is a matrix in both space and time. The space translation symmetry is broken by the entanglement cut, and the time translation symmetry is broken by choice of time $\tau_0$ for inserting the auxiliary fields in the path integral [Eq.\eqref{eq:RenyiSource}]. However, for the cylindrical subsystem geometry in Fig.\ref{supp_fig:subsys2d}, considered for the 2d Hubbard model, the  system retains the translation symmetry parallel to the ($y$) direction of the entanglement cut. In this case, one can use Fourier transform along the $y$ direction, as we discuss later. 

To solve the DMFT equations in imaginary time $\tau$ without time-translation invariance, we discretize the DMFT equations in imaginary time. We divide the time interval $[0, \beta)$ at inverse temperature $\beta=1/T$ into $N_{\tau}$ segments with discretization step $\delta \tau = \beta/N_{\tau}$. However, while discretizing we have to ensure the appropriate  anti-periodic boundary conditions on the fermionic Green's function, namely
\begin{subequations}
\begin{align}
G(\tau+\beta, \tau')& = - G(\tau, \tau') \\
G(\tau, \tau'+\beta) &= -G(\tau, \tau'), 
\end{align}
\end{subequations}
which is equivalent to the anti-periodic boundary conditions on Grassmann variables, $c_{\Nt}=-c_0$ and $\cb_{\Nt}=-\cb_0$, in the time-discretized form. Here we have suppressed the space, spin and replica indices for brevity. We use the indices $(n, m)$ running from $n,m=0$ to $\Nt$ for $(\tau,\tau')$. We also write the following useful discretization rules 
\begin{subequations}
\begin{align} 
(\partial_{\tau} \cb(\tau))c(\tau) &= \frac{(\cb_{n+1}-\cb_n)c_n}{\delta\tau},  \\
\cb(\tau)c(\tau) &= \cb_{n+1}c_n.
\end{align}
\end{subequations}
The above rules arise since the creation operator $\cb$ always appears slightly later in time than the annihilation operator $c$ in the path integral. Using the above rules, e.g., we can write Eq.\eqref{eq:compact_S_ent} as  
\beq
S_\lambda &= \delta\tau^2\sum^{\Nt}_{n,m=1} \sum_{ij,\sigma,\alpha\beta,nm}\cb_{i\sigma\alpha n}[-G^{-1}_{0, i\alpha n, j\beta m}] c_{j\sigma\beta m} \notag\\
&+U \delta\tau\sum^{\Nt}_{n=1} \sum_{i\alpha}\cb_{i\uparrow\alpha, n+1} c_{i\uparrow\alpha n}\cb_{i\downarrow\alpha, n+1} c_{i\downarrow\alpha n}. 
\eeq 
The inverse of lattice Green's function appearing above is given by
\begin{align}
-G^{-1}_{0, i\alpha n, j\beta m}&= g^{-1}_{\alpha m, \beta n}\delta_{ij} + \frac{1}{\delta\tau}t_{ij}\zeta_{mn}\delta_{\alpha\beta} \notag\\
&+ \lambda\delta_{i\in A}\delta_{ij}M_{\alpha\beta}\frac{\delta_{m, p+1}\delta_{n p}}{\delta\tau^2}
\end{align}
where the index $p\in [0, \Nt)$ is arbitrary depending on $\tau_0$. Here
\begin{subequations}
\begin{align}
g^{-1}_{\alpha m, \beta n}&=
\frac{1}{\delta\tau^2}(\zeta_{mn}-\delta_{mn})\delta_{\alpha\beta}-\mu \frac{1}{\delta\tau} \zeta_{mn}\delta_{\alpha\beta} 
\end{align}
with
\begin{align}
\zeta_{mn} &= \delta_{m, n+1} \hspace{0.4cm} n<\Nt \notag\\
&=-1 \hspace{0.4cm} n=\Nt 
\end{align}
\end{subequations}
Similarly, Eq.\eqref{eq:DMFTeq1} becomes 
\begin{align} 
\mathcal{G}_{i\alpha m, \beta n}^{-1}& = g^{-1}_{\alpha m, \beta n} - \Delta_{i\alpha m,\beta n}-\lambda\delta_{i\in A}M_{\alpha\beta}\frac{\delta_{m, p+1}\delta_{np}}{\delta\tau^2}  \label{supp_eq:impurity_G1}
\end{align}
The IPT self-energy [Eq.\eqref{eq:DMFTeq3}] is obtained as
\begin{subequations}
\begin{align} \label{supp_eq:IPT_self_E}
\Sigma_{i\alpha m, \beta n}& = U G_{i\alpha m,\beta n}\frac{\delta_{m, n-1}}{\delta\tau} - U^2 \mathcal{\tilde{G}}_{i\alpha m, \beta n}^2 \mathcal{\tilde{G}}_{i\beta m, \alpha n},
\end{align}
where
\begin{align} 
\mathcal{\tilde{G}}_{i\alpha m, \beta n} = \mathcal{G}_{i\alpha m, \beta n}-U G_{ii,\alpha m,\beta n}\frac{\delta_{m, n-1}}{\delta\tau}.
\end{align}
\end{subequations}
The lattice Green's function is obtained as 
\begin{subequations}
\begin{align} 
\sum_{n_1}\sum_{k\gamma} G^{-1}_{i\alpha m, k\gamma n_1}G_{k\gamma n_1, j\beta n}&= \delta_{ij}\delta_{\alpha\beta} \frac{\delta_{mn}}{\delta\tau^2} \label{supp_eq:latticeG_inv1}
\end{align}
where, within the local self-energy approximation, the Dyson equation is
\beq  \label{supp_eq:Ginv1}
G^{-1}_{i\alpha m, j\beta n} = G^{-1}_{0,i\alpha m, j\beta n}-\delta_{ij}\Sigma_{i\alpha m, \beta n}
\eeq 
\end{subequations}
Now one can obtain the lattice Green's function from the above equation.
The lattice Green's function determines the hybridization function through Eqs.(\ref{eq:hybridization_eq1},\ref{eq:cavity_eq1}). 

\subsection{Recursive Green's function method}\label{RecursiveG}

The most computationally expensive part of the DMFT steps discussed above is the inversion of $G^{-1}$, a matrix of dimension $\sim N_x\Nt \times N_x\Nt$, to obtain $G$ via Eq.\eqref{supp_eq:latticeG_inv1} for a lattice with $N_x$ sites along the ($x$) direction of partitioning. A direct inverse with the DMFT self-consistency loop is only feasible for relatively small systems of size $N_x\lesssim 12 $ with $\Nt \lesssim 1000$. For larger systems, we use a recursive Green's function method along the $x$ direction.  The recursive method can be implemented for the open boundary condition (OBC) as well as for the periodic boundary condition (PBC).  
In dimension $d>1$ with translation symmetry, we can make simplifications by using Fourier transform in the transverse momenta (see the discussion later). To demarcate between directions with translation symmetry and the direction of recursion $x$, we denote the lattice sites by $\bi = (i, \bi_{\perp} )$ below.

To set up the recursive method, we rewrite Eq.\eqref{supp_eq:latticeG_inv1} as a matrix equation
\begin{subequations} \label{supp_eq:latticeG_inv1_A}
\begin{align}
    \GMI \GM &= \IM \\
    \IM_{\bi\alpha m, \bj\beta n} &= \delta_{\bi\bj}\delta_{mn}\delta_{\alpha\beta},
\end{align}
where 
\begin{align}
    (\GM)_{\bi\alpha m, \bj\beta n} &= \delta\tau G_{\bi\alpha m, \bj\beta n},
\end{align}
\end{subequations}
and similarly for $\GMI$. 
We separate the system (spatially) as a system ``$S$" and the rest ``$R$", and write 
\begin{align}\label{supp:eq_1.3}
\bigg[
    \begin{pmatrix}
    (\GM_R)^{-1} & -\TM_{RS} \\
    -\TM_{RS} &  (\GM_S)^{-1}
    \end{pmatrix}
\bigg] 
    \begin{pmatrix}
    \GM_R^{R+S)} & \GM_{RS}^{(R+S)}\\
     \GM_{SR}^{(R+S)} & \GM_{S}^{(R+S)}
    \end{pmatrix} &= \IM
\end{align}
Here $\GM_R$ and $\GM_S$ are Green's functions of the system and the rest in the absence of any coupling between them. $\TM$ connects the systems and the rest, and $\GM^{(R+S)}$ is the full Green's function of the combined system. From the above, we get 
\begin{subequations}
\begin{align}
    \GM_R^{(R+S)}&=\GM_R+\GM_R\TM_{RS}\GM^{(R+S)}_{SR} \label{eq:GR_RS}\\
    \GM^{(R+S)}_{RS} &= \GM_R\TM_{RS}\GM^{(R+S)}_S \label{eq:GRS_RS}\\
    \GM^{(R+S)}_{SR} &= \GM_S\TM^{\dagger}_{RS}\GM^{(R+S)}_R \\ 
    \GM_S^{(R+S)}&=\GM_S+\GM_S\TM^{\dagger}_{RS}\GM^{(R+S)}_{RS} 
\end{align}
\end{subequations}
\subsubsection{Derivation of Eq.\eqref{eq:cavity_eq1} for the cavity Green's function}\label{RecursiveGcavity}
For this case, we take $S$ as a single site $i$ and $R$ as rest of the system. Using Eqs.(\ref{eq:GR_RS},\ref{eq:GRS_RS}), we obtain
\begin{align}
\GM_R&=\GM_R^{(R+S)}-\GM_{RS}^{(R+S)}(\GM_S^{(R+S)})^{-1}\GM_{SR}^{(R+S)}
\end{align}
The above leads to Eq.\eqref{eq:cavity_eq1} when we identify $\GM^{(i)}=\GM_R$, i.e., the cavity Green's function with $i$-th site removed, and $\GM=\GM^{(R+S)}$, the Green's function of the whole lattice.

\subsubsection{Recursive solution of Eq.\eqref{supp_eq:latticeG_inv1_A}}

We imagine successively building the system along the $x$ direction from the left, starting from the first layer at $i=l=1$ and then adding successive layers till $l=N_x$. Imagine that at the $l$-th step of recursion, we have only left $l+1$ layers, and we separate the system into left $l$ layers (``$L$"), i.e., $R$ of the preceding section, and add one layer (system ``$S$") more. We denote the Green's function of the left $l$ layers as $\GM^{(l)}$ and that of $l+1$ layers as $\GM^{(l+1)}$. From Eqs.\eqref{supp_eq:latticeG_inv1_A}
\begin{align}\label{supp:eq_1.8}
\bigg[
    \begin{pmatrix}
    (\GM^{(l)}_L)^{-1} & -\TM_{LS} \\
    -\TM^{\dagger}_{LS} &  (\GM_S)^{-1}
    \end{pmatrix}
\bigg] 
    \begin{pmatrix}
    \GM_L^{(l+1)} & \GM_{LS}^{(l+1)}\\
     \GM_{SL}^{(l+1)} & \GM_{S}^{(l+1)}
    \end{pmatrix} &= \IM.
\end{align}
where the coupling between $L$ and $S$ is given by
\begin{subequations}
\begin{align}
    \TM_{\bi,\bj} &= \boldsymbol{\zeta} t_{\bi,\bj} \hspace{0.5cm} i\leq l, j=l+1 \nonumber \\
                  &= 0 \hspace{0.5cm} \text{otherwise},
\end{align}
with
\begin{align}
     \boldsymbol{\zeta}_{\alpha n, \beta m} &= \zeta_{nm}\delta_{\alpha\beta}.
\end{align}
The inverse Green's functions of $L$ and $S$ in the absence of any coupling between them are
\begin{align}
     \bigg( \GM^{(l)}_L \bigg)^{-1}_{\bi\alpha m,\bj \beta n} &= \GM^{-1}_{\bi\alpha m,\bj\beta n} \hspace{0.5cm}i,j \leq l \\
     (\GM_S)^{-1}_{\bi\alpha m, \bj \beta n} &= \GM^{-1}_{\bi\alpha m,\bj\beta n} \hspace{0.5cm} i, j=l+1
\end{align}
\end{subequations}
Here $i\leq l$ indicates that the site $\bi$ belongs to a layer from $1$ to $l$. 
The full Green's function that we eventually want to calculate is $\GM = \GM^{(N_x)}$.  We can rewrite Eq.\eqref{supp:eq_1.8} as
\begin{subequations} \label{supp_eq:GlGlp_id1}
\begin{align} 
\GM_L^{(l+1)}&=\GM_L^{(l)}+\GM_L^{(l)}\TM_{LS}\GM^{(l+1)}_{SL}\label{supp_eq:GlGlp_id2}\\
    \GM^{(l+1)}_{LS} &= \GM^{(l)}_L\TM_{LS}\GM^{(l+1)}_S \label{supp_eq:GlGlp_id3}\\
    \GM^{l+1}_{SL} &= \GM_s \TM^{\dagger}_{SL}\GM^{(l+1)}_{L}\label{supp_eq:GlGlp_id4}\\
    \GM_S^{(l+1)}&=\GM_S+\GM_S\TM^{\dagger}_{LS}\GM^{(l+1)}_{LS} \label{supp_eq:GlGlp_id5}
\end{align}
\end{subequations}
We now rewrite Eq.\eqref{supp_eq:GlGlp_id2} keeping only the index $i$ for the layers, where all other indices $\alpha$, $m$ and $\bi_{\perp}$ are implicit in the matrices and contracted for matrix multiplications.
\begin{align}\label{supp_eq:Gij}
    \GM^{(l+1)}_{i, j} &= \GM^{(l)}_{i, j} + \sum_{i_1\leq l}\GM^{(l)}_{i, i_1}\TM_{i_1, l+1}\GM^{(l+1)}_{l+1, j} \hspace{0.4cm} i, j\leq l
\end{align}
From Eq.\eqref{supp_eq:GlGlp_id3}, we get 
\begin{align}\label{supp_eq:Gij_ilp1}
    \GM^{(l+1)}_{i, l+1} &= \sum_{i_1\leq l} \GM^{(l)}_{i, i_1}\TM_{i_1, l+1}\GM^{(l+1)}_{l+1, l+1} \hspace{0.4cm} i\leq l 
\end{align}
 Since $G^*_{i\alpha m, j\beta n}= G_{j\beta n, i\alpha m}$, we can obtain from the above 
 \begin{align}\label{supp_eq:Gij_lp1i}
\GM^{(l+1)}_{l+1, i} &= \sum_{i_1\leq l} \GM^{(l+1)}_{l+1, l+1}\TM_{l+1, i_1}\GM^{(l)}_{i_1, i} \hspace{0.4cm} i\leq l \end{align}
Using the above in Eq.\eqref{supp_eq:Gij}, we obtain for $i,j\leq l$ 
\begin{align}\label{supp_eq:Gij_final}
    \GM^{(l+1)}_{i, j} &= \GM^{(l)}_{i, j} + \sum_{i_1,i_2\leq l} \GM^{(l)}_{i, i_1}\TM_{i_1, l+1}\GM^{(l+1)}_{l+1, l+1}\TM_{l+1, i_2}\GM^{(l)}_{i_2, j}  
\end{align}
In the above equation, the only unknown quantity is $\GM^{(l+1)}_{l+1, l+1}$. This can be obtained as follows. From Eqs.(\ref{supp_eq:GlGlp_id3},\ref{supp_eq:GlGlp_id5}), we get 
\begin{align}
\GM^{(l+1)}_S &= \GM_S + \GM_S \TM^{\dagger}_{LS}\GM^{(l)}_L\TM_{LS}\GM^{(l+1)}_S   
\end{align} 
The above can be written in the form of a Dyson equation, 
\begin{align}\label{supp_eq:Dyson_eq1}
    \big( \GM^{l+1}_S\big)^{-1} &= (\GM^{(l+1)}_{l+1, l+1}\big)^{-1}=\GM_S^{-1} - \boldsymbol{\Sigma}^{(l)}
\end{align}
with the self energy
\begin{align}\label{supp_eq:selfE_dyson}
  \boldsymbol{\Sigma}^{(l)} &= \sum_{i_1, i_2 \leq l }\TM_{l+1, i_1}\GM^{(l)}_{i_1,i_2}\TM_{i_2, l+1}
\end{align}
Hence $\GM^{(l+1)}_{l+1, l+1}$ can be obtained using Eq.\eqref{supp_eq:Dyson_eq1}. Thus from Eqs.(\ref{supp_eq:Gij_ilp1}, \ref{supp_eq:Gij_lp1i}, \ref{supp_eq:Gij_final}, \ref{supp_eq:Dyson_eq1}, \ref{supp_eq:selfE_dyson}), we can construct the complete Green's function matrix $\GM^{(l+1)}_{\bi \alpha m, \bj \beta n}$ ($i, j\leq l+1$) of the system of $l+1$ layers from that of $l$ layers. The process can be applied recursively, starting with $l=1$ and continuing till $l=N_x-1$, which will yield us the Green's function of system size $N_x$, i.e., $\GM^{(N_x)}$. However, for the  DMFT self-consistency Eqs.(\ref{eq:DMFTeq1}, \ref{eq:DMFTeq2}, \ref{eq:DMFTeq3}, \ref{eq:LatticeGreenEq}, \ref{eq:hybridization_eq1}, \ref{eq:BetheApprox}), one does not need the full Green’s function at each DMFT iteration, only certain elements. In particular, if we only consider the nearest neighbor hopping, we will need to keep track of the onsite, nearest, and next-nearest neighbor Green’s functions to complete the DMFT loop using the cavity Eqs.(\ref{eq:hybridization_eq1},\ref{eq:cavity_eq1}). For the Bethe lattice approximation, nearest-neighbour sites of $i$-th site are disconnected, and from Eq.\eqref{eq:hybridization_eq1} we get
\begin{align}
\Delta_{i,\alpha\beta}(\tau,\tau')=t^2\sum_j'G^{(i)}_{j\alpha,j\beta}(\tau,\tau').
\end{align}
Here $\sum'_j$ indicates only summation over nearest neighbours of $i$. Furthermore, in the limit of large connectivity \cite{Georges}, $\GM^{(i)}=\GM$. Thus, in the large-connectivity Bethe lattice approximation, we only need to compute onsite elements of the lattice Green's function during the DMFT self-consistency loop.


 For periodic boundary condition (PBC), we need to incorporate the hopping matrix element between site $1$ and site $N_x$. We can implement this in the recursive procedure by changing hopping coupling matrix in the last iteration accordingly when we add the $l=N_x-1$ layer with single layer system $S$ to form the required system size $l+1=N_x$. In particular, we can explicitly write the hopping coupling matrix for nearest neighbour for PBC below
\begin{align}
    \TM_{\bi,\bj} &= \boldsymbol{\zeta} t_{\bi,\bj} \hspace{0.5cm} i= l; j=l+1 \hspace{0.25cm}\text{if}\hspace{0.25cm} l\leq N_x-2 \nonumber \\
    &= \boldsymbol{\zeta} t_{\bi,\bj} \hspace{0.5cm} i=1, l; j=l+1 \hspace{0.25cm}\text{if}\hspace{0.25cm} l= N_x-1 \nonumber \\
                  &= 0 \hspace{0.5cm} \text{otherwise},
\end{align}

\begin{figure}[ht!]
\centering
\includegraphics[width=0.95\linewidth]{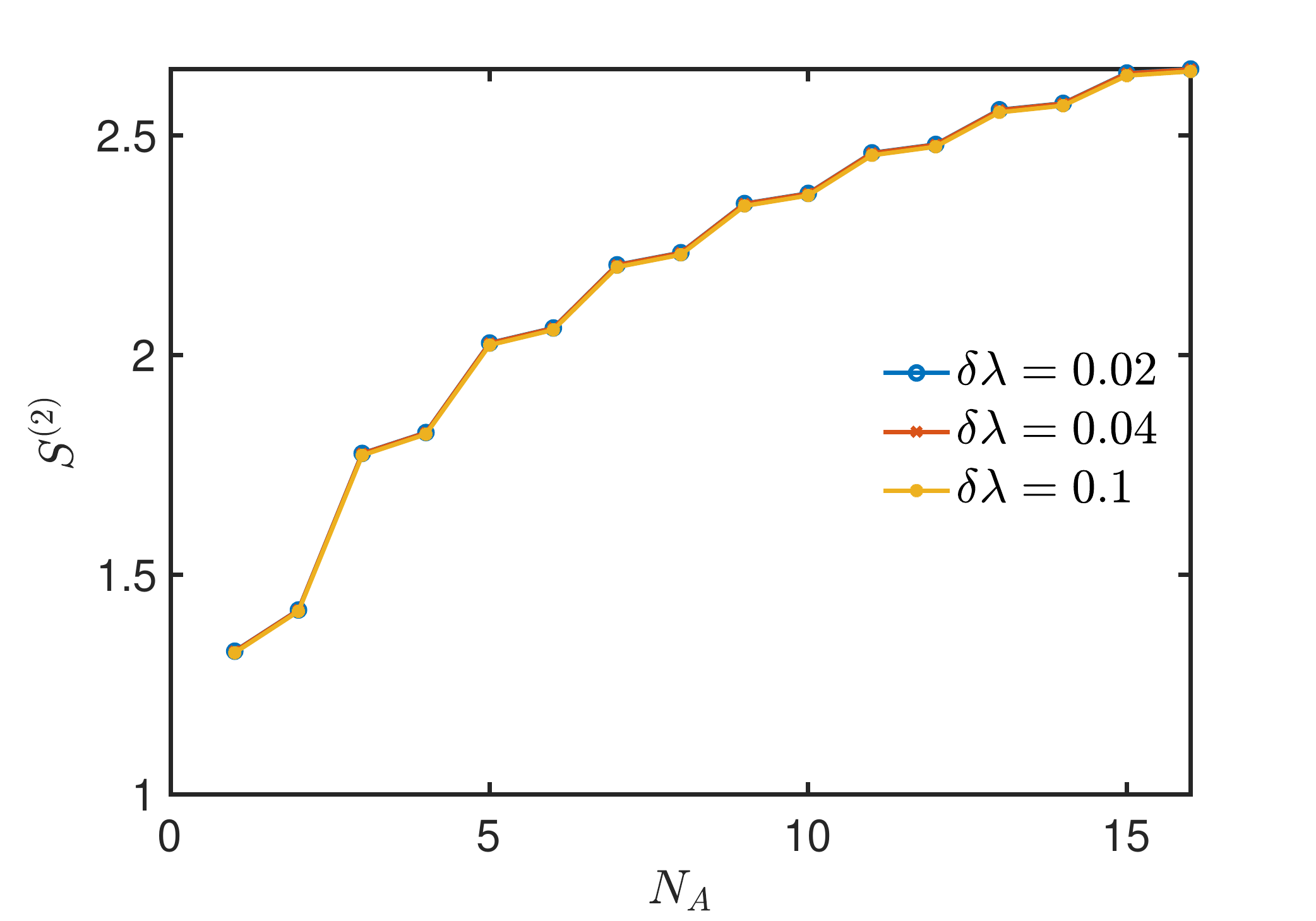}
\caption{The convergence of $\sr$ with respect to discretization $\delta\lambda$ to evaluate the integral in Eq.\eqref{eq:Avg_Skick} for $U=2.0$, $T=0.05$ and system size $N=30$ in 1d Hubbard model. 
}
\label{Supp_fig:dlamb_kind}
\end{figure} 

\section{Calculation of $\sr$ from ``kick term" integration method}\label{secsupp:kickterm}
We numerically solve the self-consistent DMFT Eqs.(\ref{supp_eq:impurity_G1}, \ref{supp_eq:IPT_self_E}, \ref{supp_eq:latticeG_inv1},\ref{eq:hybridization_eq1}) for discretized values of $\lambda \in [0, 1]$ with uniform step $\delta\lambda$ to obtain $G_{i\sigma\alpha, j\sigma\beta}(\tau_0, \tau^{+}_0)$ [Eq.\eqref{eq:RenyiGreenFn}], where we choose $\tau_0=0$. We then compute $\langle S_{\rm kick}\rangle_{Z_A^{(2)}(\lambda)}$ using Eq.\eqref{eq:Avg_Skick}.  
 To obtain $\sr$, we integrate $\langle S_{\rm kick}\rangle_{Z_A^{(2)}(\lambda)}$ over $\lambda$ from 0 to 1 using numerical interpolation over the range of $\lambda$. 

\begin{figure}[ht!]
    \centering
\includegraphics[width=0.95\linewidth]{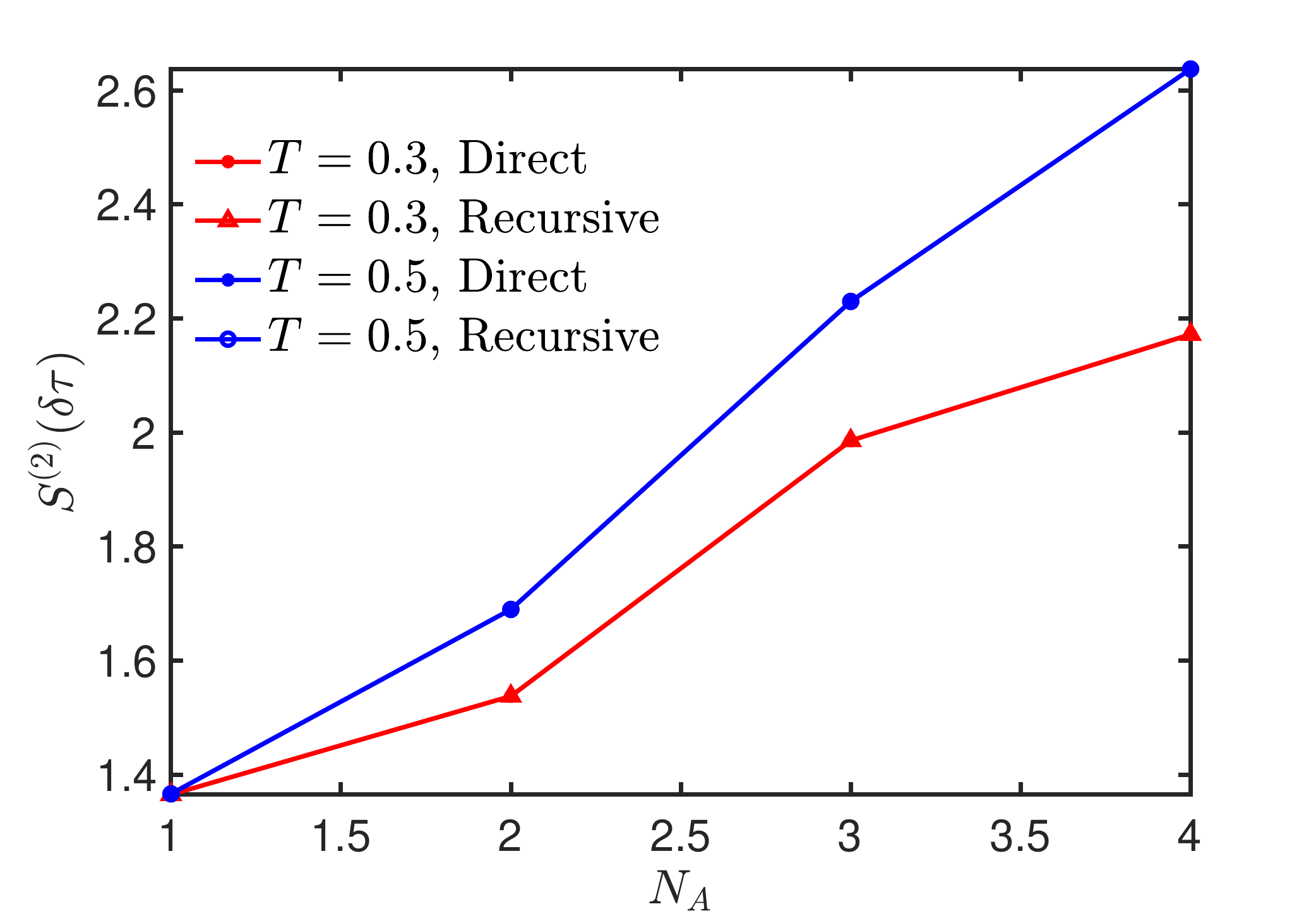}
\caption{The $\sr(\delta \tau)$ obtained using the direct and recursive inversion in the non-interacting case is shown for $T=0.3, 0.5$ and system size $N=10$. 
}
\label{supp_fig:benchmark_RG}
\end{figure} 
 
 We have used $\delta\lambda=0.1$ for most of our calculations. We have varied $\delta\lambda$ to check the convergence of $\sr$ with $\delta\lambda$. We benchmark the $\sr$ computed this way in the non-interacting case by comparing with $\sr$ obtained directly from the correlation matrix calculations, as discussed in Sec.\ref{subsec:benchmark}. For the interacting case, we have numerically checked the convergence by taking different values of $\delta\lambda$, and various numerical interpolation schemes. As an example, in Fig.\ref{Supp_fig:dlamb_kind}, we show the convergence of the $\sr$ with different $\delta\lambda=0.02, 0.04, 0.1$, for $U=2.0$, $T=0.05$, and $N=30$ in 1d. 

\section{Benchmark of the recursive Green's function method} \label{secsupp:benchmark}
 We can invert $\GM^{-1}$ in Eq.\eqref{supp_eq:latticeG_inv1} directly for small systems ($N\lesssim 12 $) to obtain the lattice Green's function. We use the recursive method for larger systems to invert $\GM^{-1}$ and obtain onsite lattice Green's function within the DMFT loop. We benchmark the recursive method by comparing computed $\sr$ with that obtained from direct inversion for $N=10$ in 1d, as shown in Fig.\ref{supp_fig:benchmark_RG} for the non-interacting case.

 \begin{figure}[ht!]
    \centering
\includegraphics[width=0.90\linewidth]{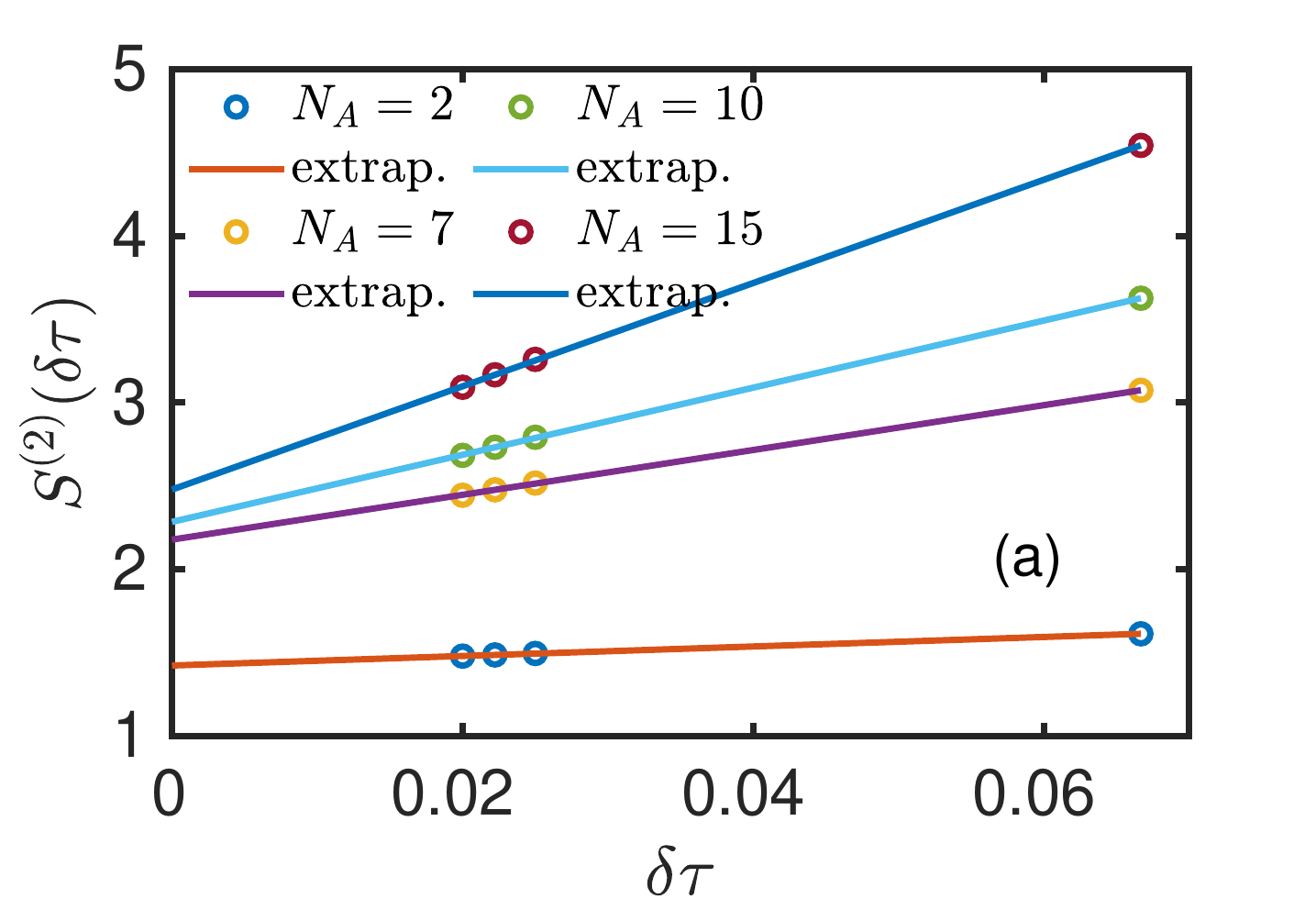}
\includegraphics[width=0.90\linewidth]{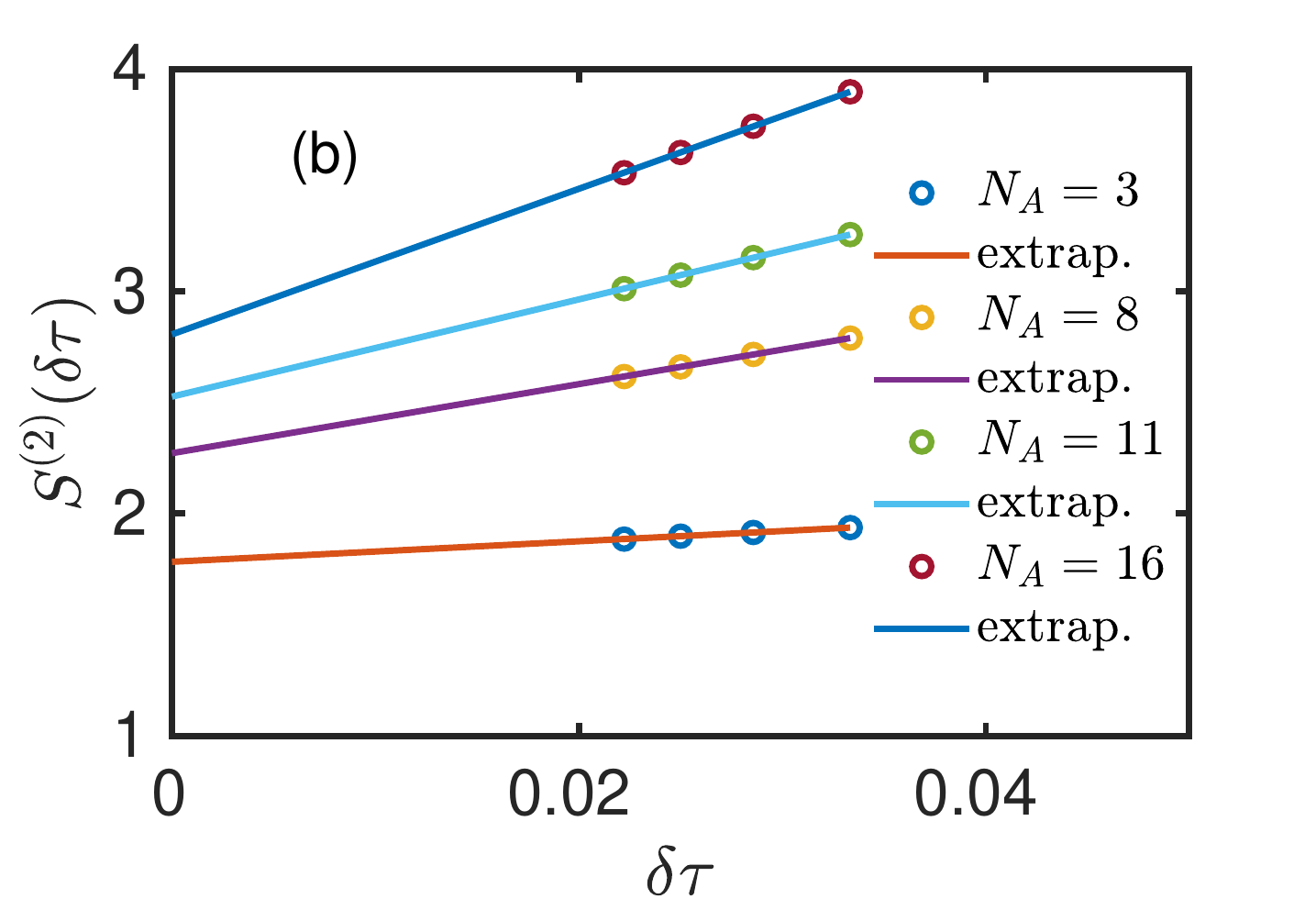}
\caption{The linear extrapolations of $\sr(\delta\tau)$ are shown here for few subsystem sizes.  The $\sr(\delta\tau)$ as a function of $\delta\tau$ and the corresponding linear extrapolation (shown as `extrap.' in legend) are shown here in (a) for $U=0.5$ and subsystem sizes $N_A=2, 7, 10, 15$ from system size $N=30$, and in (b) for $U=2$ and subsystem sizes $N_A=3, 8, 11, 16$ from system $N=50$ for at $T=0.05$. }
\label{supp_fig:lineraExtrapolation}
\end{figure} 

\section{The extrapolation of $\sr(\delta\tau)$ to $\delta\tau\to 0$ limit}\label{secsupp:extrap}
As discussed in appendix \ref{secsupp:DMFT_A}, we solve the DMFT self-consistency equations by discretizing them in imaginary time with discretization step $\delta\tau=\beta/\Nt$. 
To approach the continuum limit $\delta\tau\to 0$,  we compute $\sr(\delta\tau)$ for a few values $\delta\tau$ and then linearly extrapolate it to $\delta\tau\to 0$. We take $\delta\tau$ over the range $0.015$ to $0.075$. In most of our calculations, we take four values of $\delta\tau$ in the above range, particularly between $0.02$ to $0.04$, and then do the linear extrapolation to $\delta\tau\to 0$. It becomes progressively more challenging to take $\delta\tau$ in the above range for low temperatures $T<0.05$ as the size ($2N\Nt\times 2N\Nt$) of the Green's function matrix becomes very large. Hence we restrict our DMFT calculations up to $T=0.05$. 
We show $\sr(\delta\tau)$ as a function of $\delta\tau$ with the {linear extrapolation} in Figs.\ref{supp_fig:lineraExtrapolation}(a,b) for a few subsystem sizes for interactions $U=0.5,2$ and system size $N=30$ as an illustration of the linear extrapolation. 

\begin{figure}[ht!]
\centering
\includegraphics[width=0.90\linewidth]{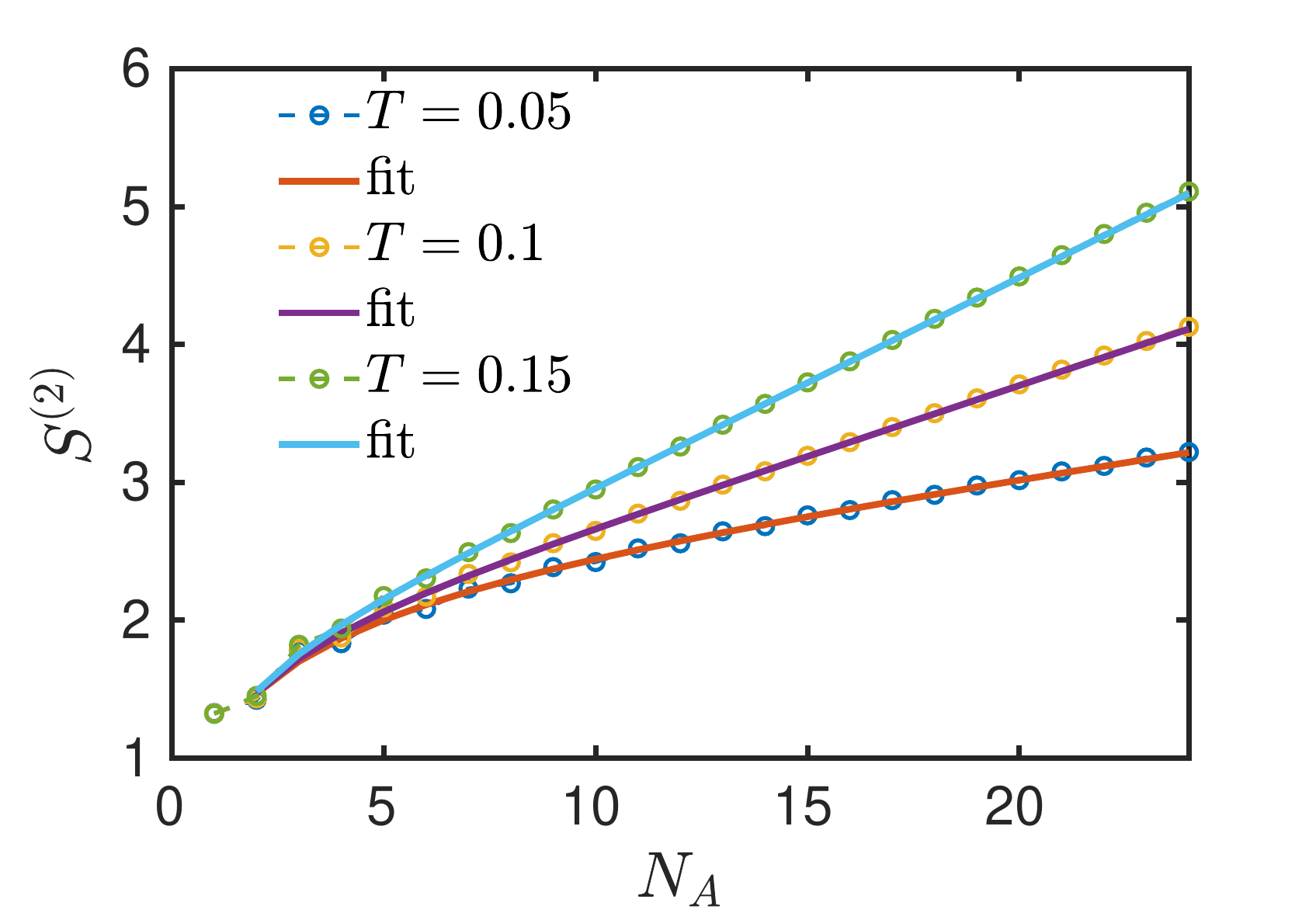}
\caption{The $\sr$ vs.~$N_A$ in 1d Hubbard model for $U=2$ and $N=50$ is shown for different temperatures with the corresponding fit to Eq.\eqref{eq:CFTscaling_3} with $c, v, b$ as free parameters.}
\label{supp_fig:fitting_cvb_1d}
\end{figure} 

\section{ Entanglement to entropy crossover in 1d Hubbard model}\label{secsupp:crossover1d}
\begin{figure}[ht!]
\centering
\includegraphics[width=0.95\linewidth]{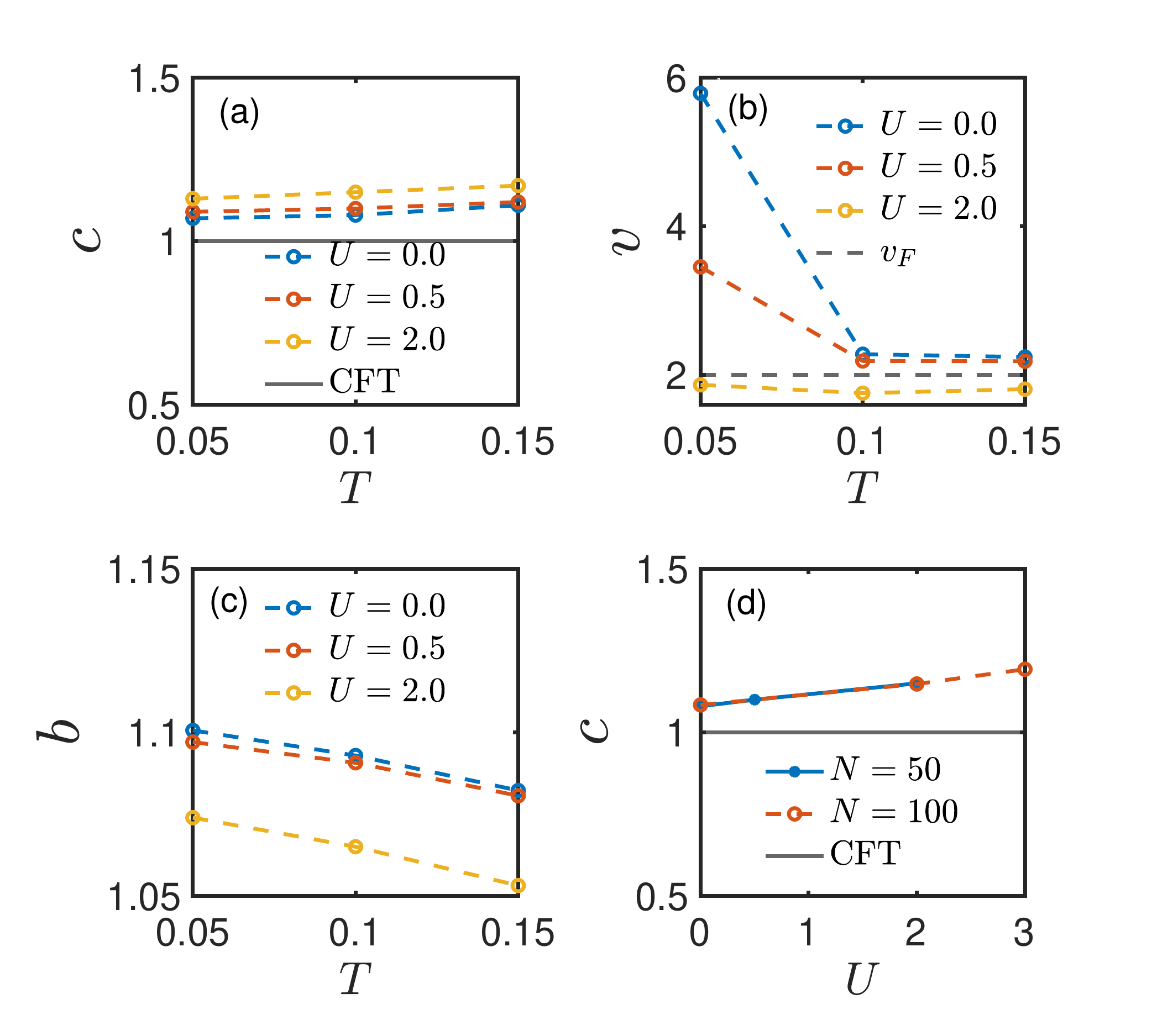}
\caption{The parameters $c, v, b$ extracted from crossover function [Eq.\eqref{eq:CFTscaling_3}] fitting, e.g., in Fig.\ref{supp_fig:fitting_cvb_1d}, are shown here for 1d Hubbard model. (a) The central charge $c$ as a function of $T$ for a few $U$ is shown and compared with the CFT value $c=1$. (b) The renormalized Fermi velocity $v$ as a function of $T$ is shown for different $U$, and compared with the non-interacting value $v_F$.  
(c) The non-universal parameter $b$ as a function of $T$.
(d) System size dependence of extracted $c$ vs. $U$ at $T=0.1$.
}
\label{supp_fig:Summary1d_cvb}
\end{figure}

Here we discuss the fitting of $\sr$ in 1d Hubbard model with the crossover function of Eq.\eqref{eq:CFTscaling_3}. 
 We first fit $\sr(N_A,T)$ with Eq.\eqref{eq:CFTscaling_3} by varying all the parameters $c, v, b$, as shown in Fig.\eqref{supp_fig:fitting_cvb_1d}. The variations of the extracted fitting parameters $c, v, b$ with temperature are shown in Fig.\eqref{supp_fig:Summary1d_cvb}. As evident from the figure, the expected CFT crossover formula [Eq.\eqref{eq:CFTscaling_3}] describes $\sr$ for the DMFT metallic state of 1d Hubbard model quite well. The extracted central charge slowly approaches the CFT value $c=1$ with decreasing temperature [Fig.\eqref{supp_fig:Summary1d_cvb}(a)] and converges well with system size $N$ [Fig.\eqref{supp_fig:Summary1d_cvb}(d)].

Here we have treated $c,v,b$ in Eq.\eqref{eq:CFTscaling_3} as fitting parameters to describe $\sr(N_A,T)$ in the 1d Hubbard model. In the next section, we first fix the ratio $(c/v)$ using the specific heat calculated from equilibrium DMFT and then fit our results for $\sr$ with Eq.\eqref{eq:CFTscaling_3} treating only $c$ and $b$ as fitting parameters. 

\subsection{Calculation of the ratio $(c/v)$ from equilibrium DMFT}\label{secsupp:cv_eqbDMFT}
At low temperature ($T\to 0$), the specific heat $c_V$ can be obtained from CFT \cite{Korepin,Cardy} as 
\begin{align}
    c_V = \frac{\pi T}{3 }\left(\frac{c}{v}\right).\label{eq:CFTcV}
\end{align}
We compute the specific heat $c_V$ for the paramagnetic metallic state in 1d Hubbard model from equilibrium DMFT calculation, as described below. 

In equilibrium, due to time translation symmetry, $G_{ij}(\tau,\tau')=G_{ij}(\tau-\tau')$. Thus we can write the DMFT self-consistency equations \cite{Georges} as follows
\begin{align}
    \mathcal{G}_i^{-1}(\tau-\tau') &= -(\partial_{\tau}-\mu)\delta(\tau-\tau')-\Delta_i (\tau-\tau') \\
    G^{-1}_i(\tau) &= \mathcal{G}_i^{-1}(\tau)-\Sigma_i(\tau)
\end{align}
where $\mathcal{G}_i(\tau)$ and $G_i(\tau)$ are the bare and full impurity Green's functions at site $i$. Furthermore, since the model [Eq.\eqref{eq:HubbardModel}] is space translation invariant, all the sites are equivalent, unlike for the DMFT in the presence of entanglement cut in Sec.\ref{sec:DMFTEntanglement}. 

\begin{figure}[ht!]
\centering
\includegraphics[width=0.90\linewidth]{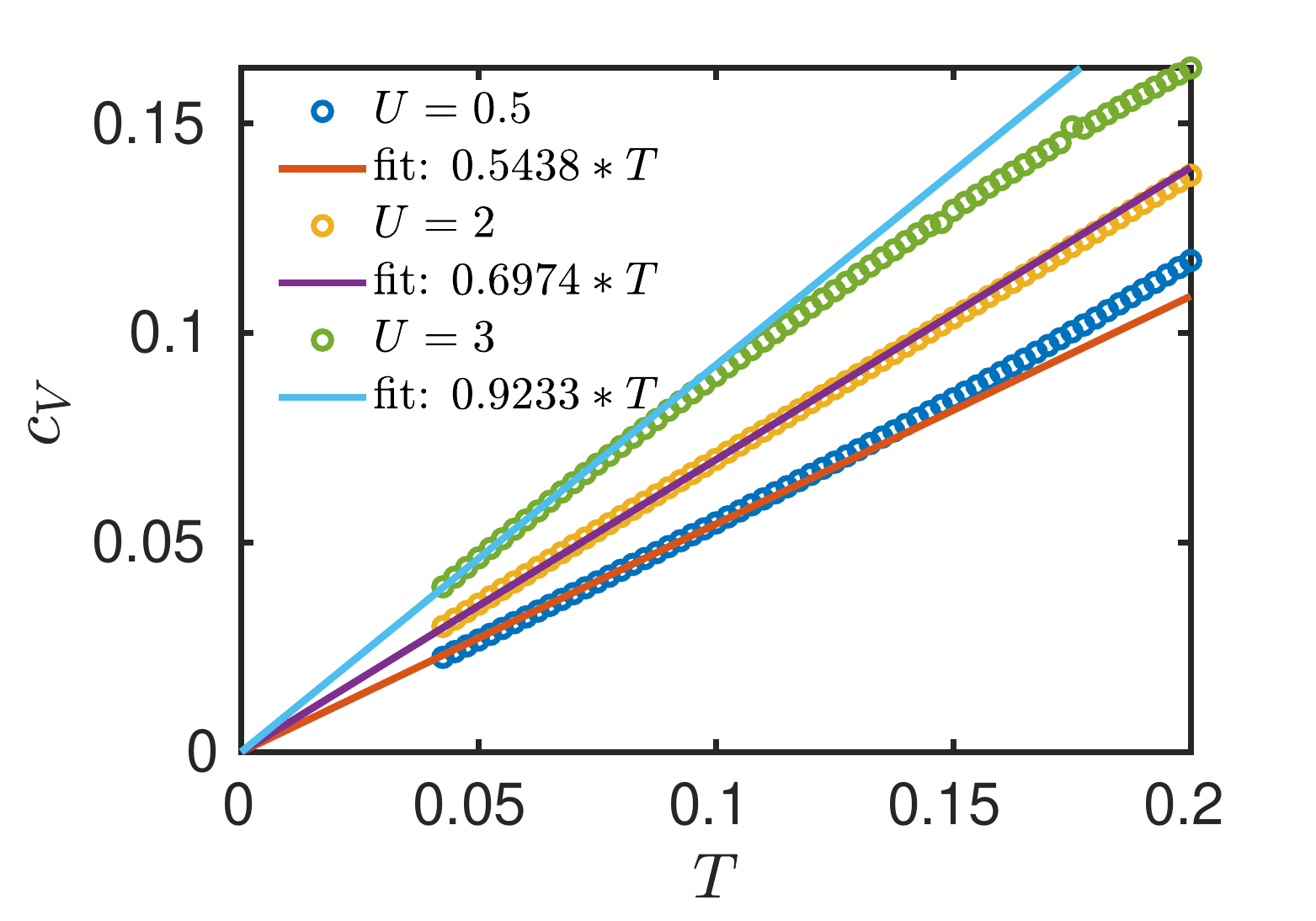}
\caption{The specific heat ($c_V$) as a function of temperature $T$ for 1d Hubbard model is shown for different interaction $U=0.5, 2, 3$. The low-temperature linear fit to $c_V(T)\propto T$ is shown and the extracted slopes are indicated in the legends. }
\label{supp_fig:cv_vs_T}
\end{figure} 

The hybridization function is given by 
\begin{align}
    \Delta_i (\tau) = \sum_{jl} t_{ij}t_{il} G^{(i)}_{jl}(\tau) 
\end{align}
We use the large-connectivity Bethe lattice approximation for the cavity Green's function, i.e., $G^{(i)}_{jl}=G_{jl}$, and $G^{(i)}_{jl}\simeq \delta_{jl}G_{jj}$. Therefore the hybridization function for nearest-neighbor hopping becomes 
\begin{align}
    \Delta_i(\tau) = zt^2 G(\tau)
\end{align}
where $G(\tau)$ is the onsite lattice Green's function and $z=2d$ is the coordination number for $D$-dimensional hypercubic lattice. The lattice Green's function is obtained using the local self-energy approximation, i.e., the lattice self-energy $\Sigma_{ij}$ is replaced by the impurity self-energy, $\Sigma_{ij}=\Sigma\delta_{ij}$, so that 
\begin{align}
    G(i\omega_m) = \int d\epsilon \frac{g(\epsilon)}{\imath\omega_m+\mu-\epsilon-\Sigma(\imath\omega_m)} 
\end{align}
where $\omega_m=(2m+1)\pi/\beta$, with $m$ an integer, is the fermionic Matsubara frequency, and $g(\epsilon)$ is the non-interacting density of states per site. The self-energy within IPT approximation \cite{Georges} is given by
\begin{align}
    \Sigma(\tau) = Un - U^2 \mathcal{\tilde{G}}^2(\tau)\mathcal{\tilde{G}}(-\tau) \\
    \mathcal{\tilde{G}}^{-1}(\tau) = \mathcal{G}^{-1}(\tau)-Un
\end{align}
where $n$ is the occupation number of a site; $n=1/2$ at half filling. For calculating thermodynamic properties such as specific heat, we solve the above equilibrium DMFT self-consistency equation following standard procedure \cite{Georges} to obtain $G(\imath\omega_m)$ and $\Sigma(\imath\omega_m)$.

Using these, we compute the internal energy $E$ \cite{Georges} from 
\begin{align}
    \frac{E}{N}& = 2T\sum_{m}\int^{\infty}_{-\infty} d\epsilon \frac{\epsilon D(\epsilon)}{i\omega_m+\mu-\Sigma_(i\omega_m) -\epsilon} \nonumber \\
    &+  T\sum_{m}\Sigma(i\omega_m) G(i\omega_m).
\end{align}
 The specific heat per site is obtained by taking the numerical derivative of internal energy, i.e., $c_{V}=(1/N)\big(\partial E/\partial T\big)$. 

\begin{figure}[ht!]
    \centering
\includegraphics[width=0.90\linewidth]{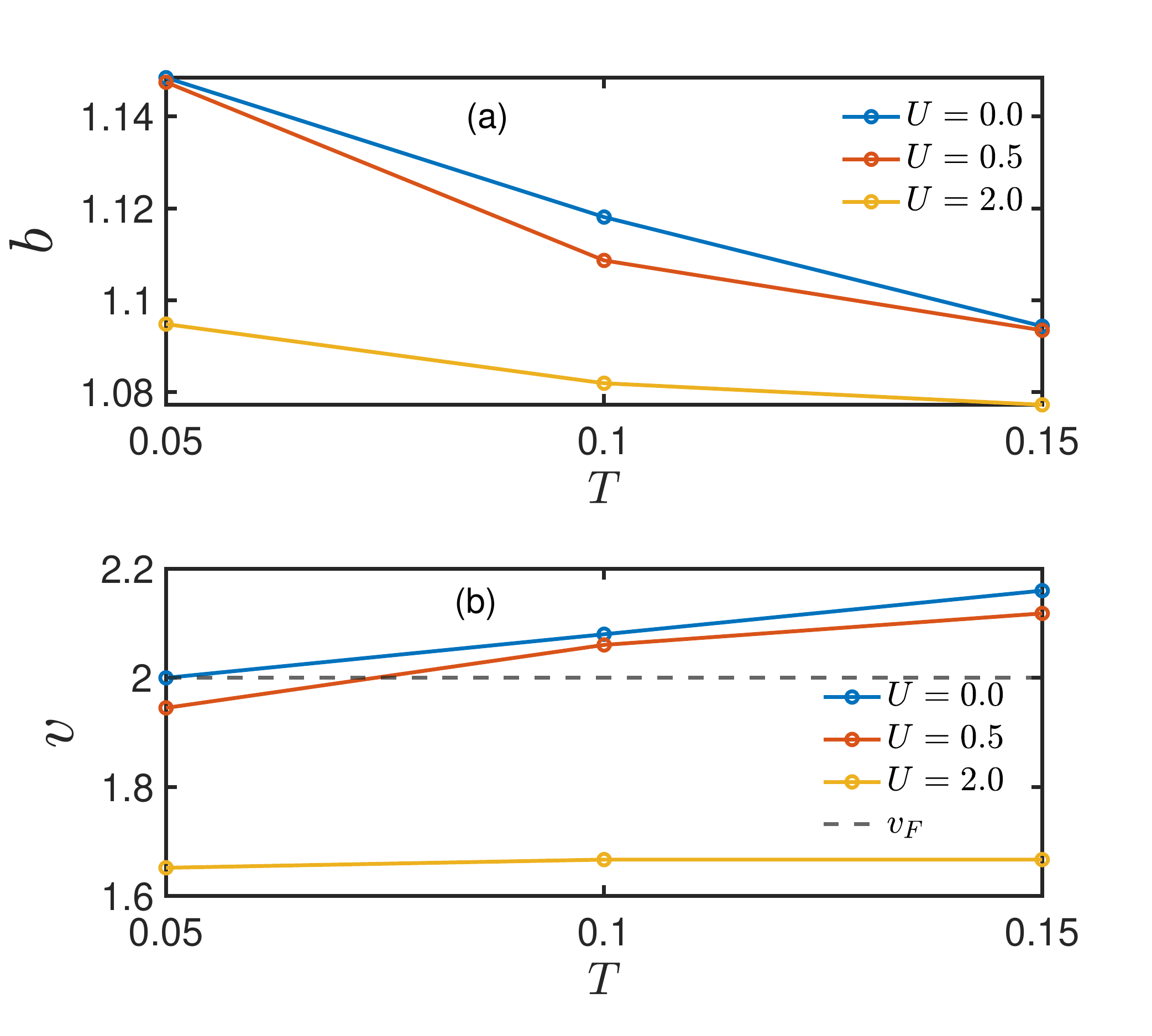}
\caption{Temperature dependence of (a) $b$ and (b) $v$ extracted by fixing $(c/v)$ ratio from specific heat and fitting CFT crossover formula [Eq.\eqref{eq:CFTscaling_3}] to the DMFT results for $\sr(N_A,T)$ in the 1d Hubbard model for $U=0.0, 0.5,2$ and system size $N=50$. The non-interacting Fermi velocity $v_F$ is shown in plot (b) for comparision with renormalised velocity $v$. }
\label{supp_fig:nonuniversal_b}
\end{figure} 
 
 We calculate the specific heat as a function of temperature for 1d and 2d Hubbard models. We show $c_V$ vs.~$T$ in 1d Hubbard model in Fig.\ref{supp_fig:cv_vs_T}. We extract $(c/v)$ from the slope of the  linear fit to $c_V(T)$ at low temperature using Eq.\eqref{eq:CFTcV}, as shown in the figure. For the half-filled 2d square-lattice Hubbard model with nearest-neighbor hopping, due to the van Hove singularity of the non-interacting band at the Fermi energy $c_V$ has $\ln T$ correction to Eq.\eqref{eq:CFTcV}, and the $(c/v)$ ratio cannot be estimated reliably. Thus we only use the $(c/v)$ ratio in 1d to fit $\sr(N_A,T)$ with the crossover formula [Eq.\eqref{eq:CFTscaling_3}], as discussed in Sec.\ref{sec:1DHubbard}. The central charge $c$ extracted this way is shown in Fig.\ref{mainfig_c_sumarry_1d}. We show the non-universal constant $b$ and the velocity $v$, obtained using the $(c/v)$ ratio and $c$, in Figs.\ref{supp_fig:nonuniversal_b}(a,b). As evident, $v$ extracted this way for weak interaction matches quite well at low temperature with non-interacting $v_\mathrm{F}$, unlike the $v$ extracted by fitting the CFT formula with three parameters $c,~v,~b$ in Fig.\ref{supp_fig:Summary1d_cvb}(b). For the 2d Hubbard model, we extract $c$ by fitting the crossover formula [Eq.\eqref{eq:CFTscaling_3}] to the computed $\sr(N_A,T)$ using $c,v,b$ as free parameters, as discussed in Sec.\ref{sec:2DHubbard}. 

\section{$\sr(N_A,T)$ for open boundary condition (OBC) in 1d Hubbard model}\label{secsupp:OBC}
\begin{figure}[ht!]
\centering
\includegraphics[width=0.90\linewidth]{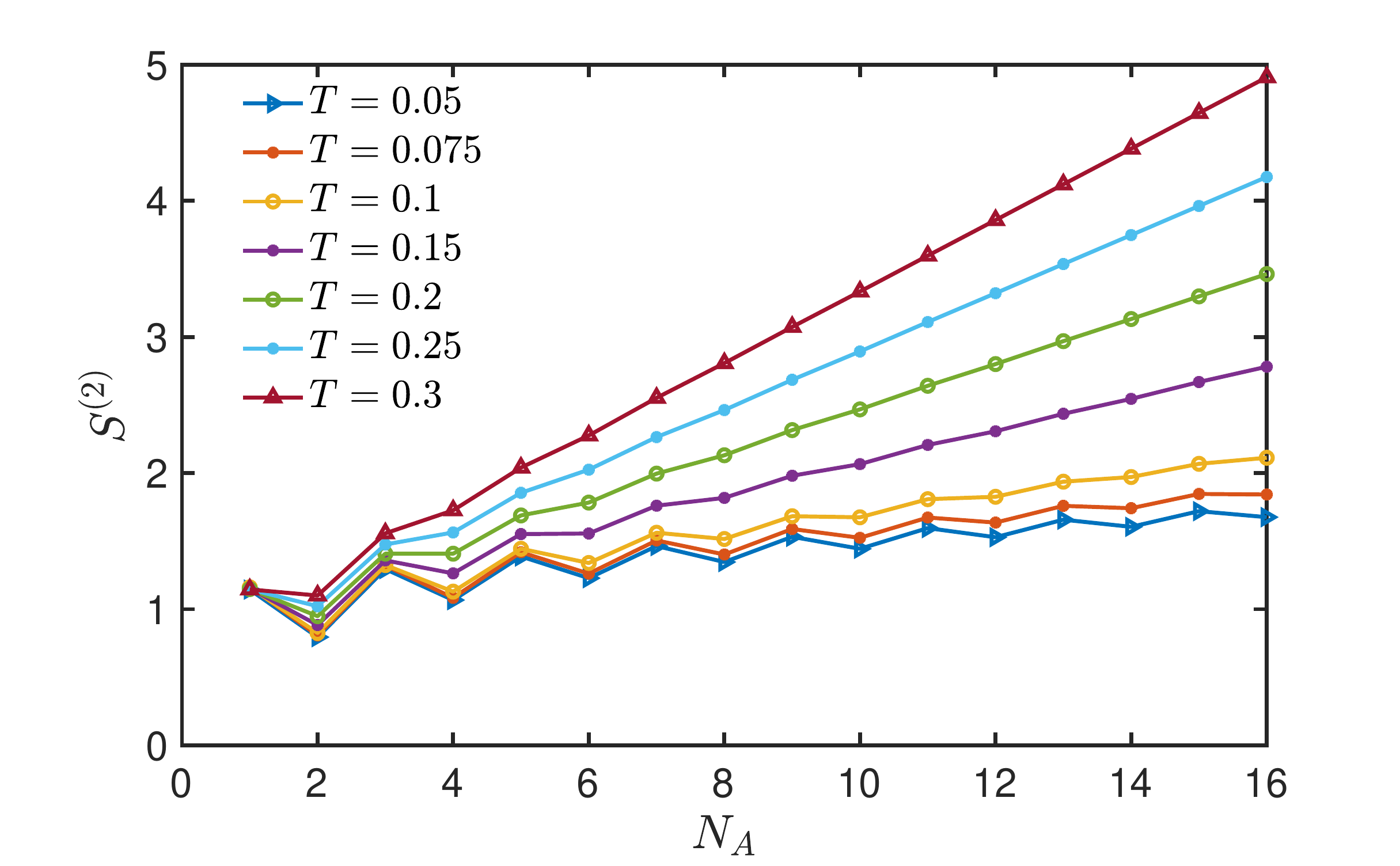}
\caption{The $\sr$ as a function of $N_A$ is shown here for different $T$ for open boundary condition (OBC) for $U=2.0$. The system size is $N=30$ here. 
}
\label{supp_fig:OBC}
\end{figure} 

We show the second R\'{e}nyi entropy $\sr$ computed via DMFT for open boundary condition (OBC) in the 1d Hubbard model for system size $N=30$ in Fig.\ref{supp_fig:OBC}. We see quite large oscillations in $\sr$ between the odd and even subsystem sizes at low temperatures for OBC. Such oscillations are present for periodic boundary condition also, e.g., in Fig.\ref{mainfig_result_1d}, but are much weaker. These oscillations, with frequency $2k_\mathrm{F}$ determined by the Fermi wave vector $k_\mathrm{F}$, are expected\cite{Calabrese2010,Swingle2013} due to the subleading corrections to the CFT result [Eq.\eqref{eq:2dcrossover_finalform}], and appear to be enhanced in OBC compared to that in PBC.

\section{DMFT equations for computing second R\'{e}nyi entropy in 2d}\label{secsupp:DMFT_2d}

In two dimensions (2d), we take a cylindrical geometry for the subsystem $A$ to compute the second R\'{e}nyi entropy, as shown in Fig.\ref{supp_fig:subsys2d}. 
We take the entanglement cut parallel to the $y$ axis, i.e., partition the system along the $x$ direction. The most difficult part in solving the non-equilibrium DMFT Eqs.(\ref{eq:DMFTeq1}, \ref{eq:DMFTeq2}, \ref{eq:DMFTeq3}, \ref{eq:LatticeGreenEq}, \ref{eq:hybridization_eq1}, \ref{eq:BetheApprox}) is the inversion of the inverse lattice Green's function. We rewrite Eq.\eqref{eq:LatticeGreenEq} below as 
\begin{align} \label{eq:LatticeGreenEq2d}
\int_{0}^{\beta} & d\tau'' \sum_{\br''\gamma} \big[G^{-1}_{0, \br\alpha, \br''\gamma}(\tau, \tau'')-\delta_{\br,\br''}\Sigma_{\br,\alpha\gamma}(\tau,\tau'')] \nonumber \\
&G_{\br''\gamma,\br'\beta}(\tau'', \tau') 
= \delta_{\br,\br'}\delta_{\alpha\beta}\delta(\tau-\tau') 
\end{align}
where $\br=(x,y)$ represents two dimensional co-ordinates. Due to translation symmetry in the $y$ direction, $G_{\br\alpha,\br'\beta}(\tau,\tau')=G_{x\alpha,x'\beta, y-y'}(\tau,\tau')$. As a result, Green's function can be represented using Fourier transform along the $y$ direction with momentum $k_y$, 
\begin{align}\label{sup_eq:Gabxy}
    G_{x\alpha,x'\beta, y-y'}(\tau,\tau') = \frac{1}{N_y} \sum_{k_y} e^{-ik_y(y-y')} G_{x\alpha,x'\beta, k_y}(\tau,\tau').
\end{align}
Thus, from Eq.\eqref{eq:LatticeGreenEq2d}, we can write 
\begin{align} \label{eq:LatticeGreenEq2dky}
\int_{0}^{\beta} & d\tau'' \sum_{x''\gamma} \big[G^{-1}_{0, x\alpha, x''\gamma, k_y}(\tau, \tau'')-\delta_{xx''}\Sigma_{x,\alpha\gamma}(\tau,\tau'')] \nonumber \\
&G_{x''\gamma,x'\beta, k_y}(\tau'', \tau') 
= \delta_{xx'}\delta_{\alpha\beta}\delta(\tau-\tau') 
\end{align}
for each $k_y$ mode, where 
\begin{align}
    G^{-1}_{0, x\alpha, x'\gamma, k_y}(\tau, \tau')=& \big[ (-\partial_{\tau}+\mu-2t\cos{k_y})\delta_{xx'}\big.\nonumber \\
    &\big.-t_{xx'}\big]\delta_{\alpha\beta}\delta(\tau-\tau')
\end{align}
where $t_{xx'}$ is hopping amplitude along $x$ direction. For nearest-neighbor hopping, we have $t_{x, x\pm 1}=t$, and $t_{xx'}=0$ otherwise. 
The hybridization function with large-connectivity Bethe lattice approximation is given by
\begin{align}
    \Delta_{x,\alpha\beta} &= t^2\bigg[ 
    G_{x-1,\alpha, x-1,\beta, y-y'=0}  +   G_{x+1,\alpha, x+1,\beta, y-y'=0} \bigg. \nonumber \\
    &\bigg.+ 2  G_{x\alpha, x\beta, y-y'=0}\bigg]. 
\end{align}
In the above equation, we have omitted the time arguments $(\tau,\tau')$ for notational convenience. The Green's function $G_{x\alpha, x\beta, y-y'=0}(\tau,\tau')$ is obtained from Eq.\eqref{sup_eq:Gabxy} as 
\begin{align}
    G_{x\alpha, x\beta, y-y'=0}(\tau,\tau') = \frac{1}{N_y}\sum_{k_y} G_{x\alpha, x\beta, k_y}(\tau,\tau') 
\end{align}
We use the recursive Green's function method in the $x$ direction for $G_{x\alpha, x\beta, k_y}(\tau,\tau')$, as described in Appendix \ref{secsupp:DMFT_A}, to obtain the lattice Green's function for each $k_y$ mode. 

\section{Widom formula for $S^{(2)}_A(N_A, T)$ in 2d}\label{secsupp:widomformula}

\begin{figure}[ht!]
    \centering
\includegraphics[width=0.80\linewidth]{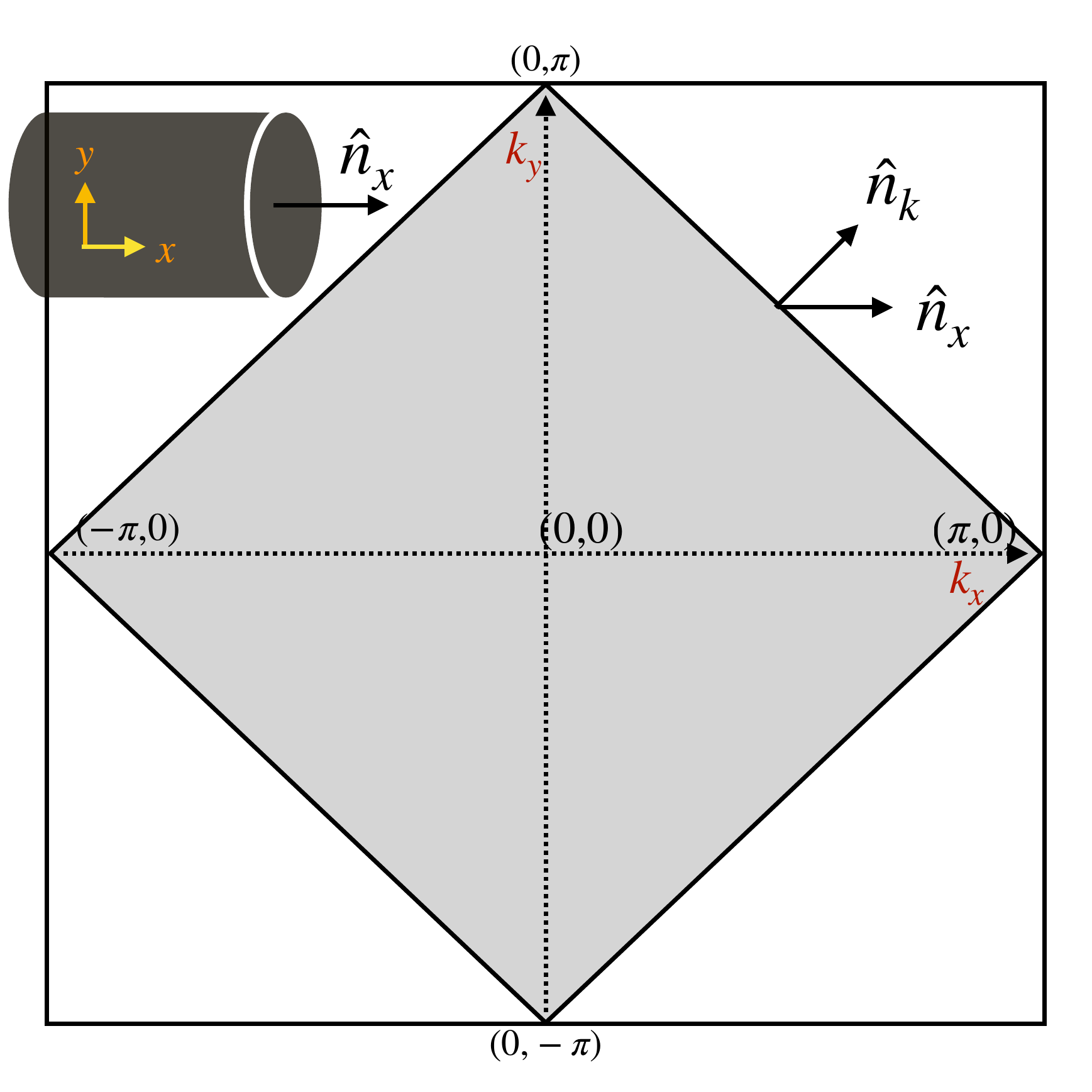}
\caption{ A schematic of the Fermi sea at half-filling (in gray color) for non-interacting 2d tight binding model. The real-space subsystem of cylindrical geometry is shown in the top left corner. $\hat{n}_k$ is a unit vector perpendicular to Fermi-surface and $\hat{n}_x$ is perpendicular to real-space boundaries of the subsystem.}
\label{supp_fig:FermiSurface}
\end{figure} 

As we discussed in the main text, in the Widom formula, the effective number of modes from Fermi surface assuming made of independent patches is given by Eqn.\ref{eq:Nmodes} and we rewrite it here 
\begin{align}\label{eq_supp:Nmodes}
N_{\rm modes} = \frac{1}{(2\pi)^{d-1}}\frac{1}{2}\int_{\partial A_x}\int_{\partial A_k} dA_x dA_k |\hat{\boldsymbol{n}}_x\cdot \hat{\boldsymbol{n}}_k| 
\end{align}
We use the 2d non-interacting dispersion $\varepsilon_k=-2t\cos k_x - 2t\cos k_y$ for nearest neighbour hopping. The Fermi-surface at half filling is shown in Fig.\ref{supp_fig:FermiSurface}.   The unit normal to Fermi-surface $\hat{\boldsymbol{n}}_k=\frac{1}{\sqrt{2}}(\pm \hat{x}\pm \hat{y})$.  For cylindrical geometry, we have two interface parallel to $y$-axis and hence the unit normal to real space $\hat{\boldsymbol{n}}_x=\pm \hat{x}$. Hence, we get $|\hat{\boldsymbol{n}}_k\cdot \hat{\boldsymbol{n}}_x| = \frac{1}{\sqrt{2}}$.
Therefore, 
\begin{align}
    N_{\rm modes} = \frac{1}{4\pi}\times 2N_y \times (4\sqrt{2}\pi ) \times \frac{1}{\sqrt{2}} = 2N_y
\end{align}
where the contribution $2N_y$ comes from integration over real space boundary and $4\sqrt{2}\pi$ comes from integration over the Fermi-surface.   

\begin{figure}[ht!]
    \centering
\includegraphics[width=0.90\linewidth]{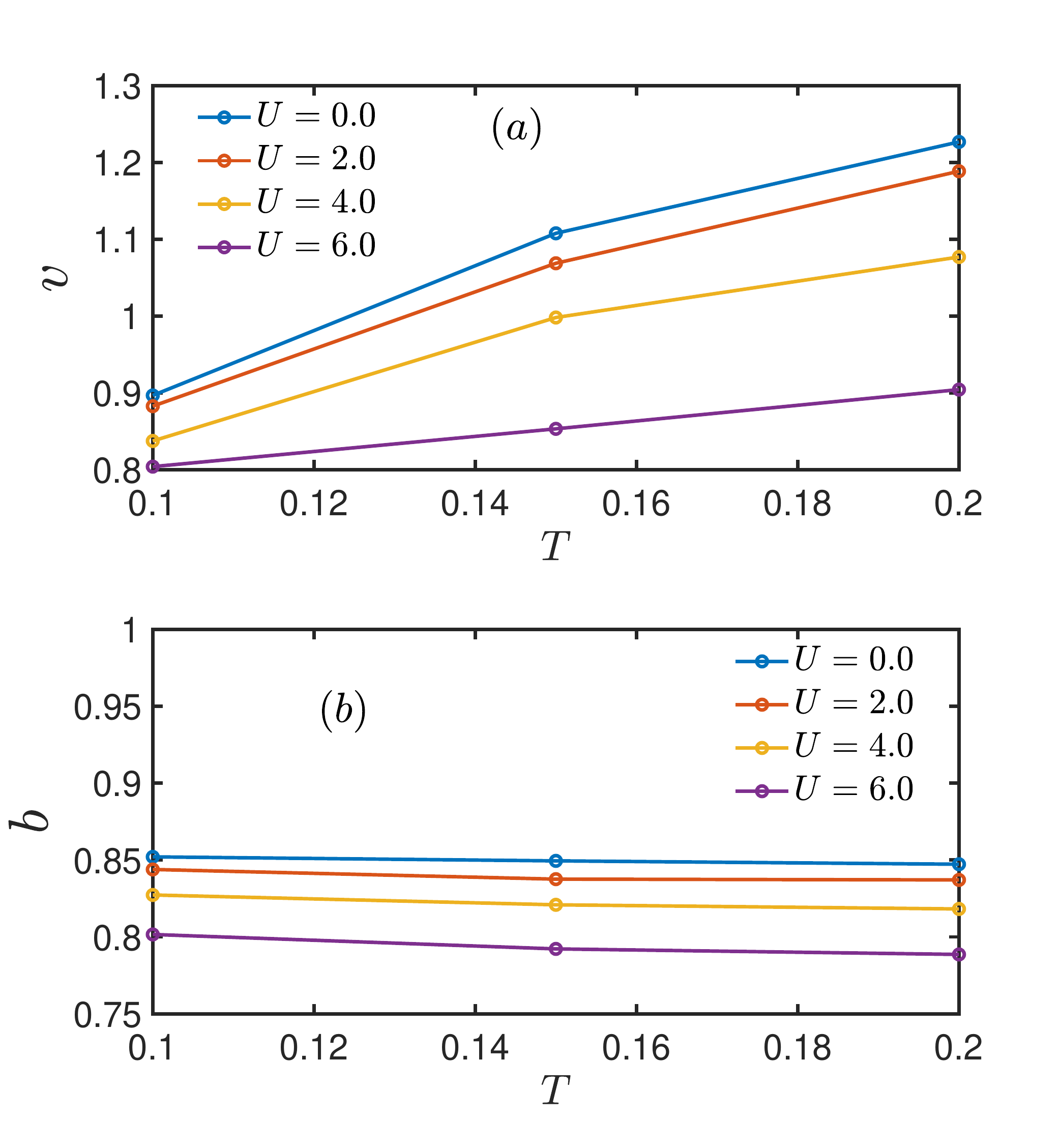}
\caption{The temperature dependence of (a) $v$, and (b) $b$, extracted by fitting Widom crossover formula  of Eq.\eqref{eq:2dcrossover_finalform} to $S^{(2)}/N_y$ for different $U$ computed from system size $20\times 20$. }
\label{supp_fig:Fitpara2d}
\end{figure}

We fit the numerically computed $S^{(2)}/N_y$ data within DMFT to the Widom crossover formula Eq.\eqref{eq:2dcrossover_finalform} and the extracted central charge $c$ as a function of $T$, $U$ and system size are shown in the main text. 
In the Fig.\ref{supp_fig:Fitpara2d}, the temperature dependence of renormalized velocity $v$ and non-universal constant $b$ extracted by fitting to Eq.\eqref{eq:2dcrossover_finalform} are shown for different interaction $U$. This parameters ($v, b$) are shown for system $20\times 20$. 

 \clearpage
\bibliography{RenyiEntropyInHubbardModelDMFTapprox}

\end{document}